\documentclass[structabstract]{aa}  
\usepackage{graphicx}
\usepackage[authoryear]{natbib}
\bibpunct{(}{)}{;}{a}{}{,} 

\begin{document}
   \title{Probing the anomalous extinction of four young star clusters: the use of colour-excess, main sequence fitting and fractal analysis.}

   \titlerunning{Probing the anomalous extinction of young star clusters}
\authorrunning{Fernandes et al.}

   \author{B. Fernandes
          \inst{1}
          \and
          J. Gregorio-Hetem
          \inst{1}
          \and
          A. Hetem Jr.
          \inst{2}
          }

   \institute{Universidade de S\~ao Paulo, IAG, Rua do Mat\~ao 1226, 05508-900 S\~ao Paulo, Brasil\\
              \email{bfernandes@astro.iag.usp.br}
         \and
           Universidade Federal do ABC, CECS, Rua Santa Ad\'elia, 166, 09210-170 Santo Andr\'e, SP, Brazil\\
           }

 
  \abstract
  {} 
   {Four  young star clusters were studied in order to characterize their anomalous extinction or variable reddening that
could be due to a possible contamination by dense clouds or circumstellar effects.}
   {The extinction law ($R_V$) was evaluated by adopting two methods: (i) the use of theoretical expressions based on the colour-excess 
of stars with known spectral type, and (ii) the analysis of two-colour diagrams, where the slope of observed colours distribution
is compared to the normal distribution. An algorithm to reproduce the  zero age main sequence (ZAMS) reddened colours was developed in 
order to derive the average visual extinction ($A_V$) that provides the best fitting of the observational data. The structure of the clouds was evaluated
by means of statistical fractal analysis, aiming to compare their geometric structure with the spatial distribution of the cluster members.}
   {The cluster NGC~6530 is the only object of our sample showing anomalous extinction. In average, the other clusters are
suffering normal extinction, but several of their members, mainly in NGC~2264, seem to have high $R_V$, probably due to 
circumstellar effects. The ZAMS fitting provides $A_V$ values that are in good agreement with those found in the literature.  The fractal analysis 
shows that NGC~6530 has a centrally concentrated distribution of stars that is different of the sub-structures found in the density distribution of  
the cloud projected in the $A_V$ map, suggesting that the original cloud has been changed with the cluster formation. 
On the other hand, the fractal dimension and the statistical
parameters of Berkeley~86, NGC~2244, and NGC~2264 indicate a good cloud-cluster correlation, when compared to other works based on artificial distribution of 
points.}
   {}

  
   \maketitle
   
   %

\section{Introduction}

Extinction is one of the best known consequences of the dust present in the Interstellar
Medium (ISM). The study of extinction on star-forming regions reveals properties of the interstellar material
that leads to a better understanding of the formation and evolution of stars.

Each line-of-sight has an extinction law, which is characterized by the ratio of the total-to-selective extinction $R_V = A_V$ / E(B-V) and depends on the composition of the ISM. According to \cite{1979ARA&A..17...73S}  its average value in the diffuse ISM is 3.1, while in denser regions it might reach values in the range 4 $< R_V<$ 6.  
Anomalous high  extinction laws have been found in several  star-forming regions   \citep{1981A&AS...45..451N, ck83, 1990A&A...227..213C, 2000PASJ...52..847P, 2007ApJ...671..555S}. Besides the differences of the ISM characteristics, high values of $R_V$  are commonly explained by selective evaporation of small grains caused by the radiation from hot stars, or by grain growth in circumstellar environments (e.g. van den Ancker et al. 1997). The circumstellar hypothesis seems to be supported by the correlation between reddening
and extinction found by Sung et al. (2000) for a sample of 30 stars. Most of these objects had their SED well fitted by assuming $R_V$ = 3.1, with some exceptions that show anomalous  $R_V$ and have high extinction probably due to the circumstellar material. From the results of a polarimetric survey in the
direction of the Lagoon Nebula (M8), \citet{mccall90} estimated  $R_V$ = 4.64 $\pm$ 0.27 that is attributed  to circumstellar effects. This anomalous high $R_V$ was
obtained after removing the normal foreground extinction. 

The dependence of the extinction law as a function of wavelength has also been widely discussed in the literature, given the interest in identifying the characteristics
 of different types of grains causing the extinction, as well as the need to properly correct the reddening that affects the observational data. 
In general a ``universal extinction law", usually a power law for wavelengths larger than 1$\mu$m, has been adopted. The power law ($A_{\lambda} \propto \lambda^{-\beta}$) 
fits most extinction curves. However, it is recognized that the index $\beta$ varies significantly depending on the line-of-sight.
Based on data from the {\it Hubble Space Telescope}, \citet{2009ApJ...699.1209F} investigated the extinction dependence on near-infrared (near-IR) wavelengths. They found that a ``universal" law does not apply in this case and the index $\beta$ has a tendency to decrease with increasing $R_V$.  Instead of extrapolating the usual 
power law, \cite{2009ApJ...699.1209F} suggest a different form, independent of wavelength, that better describes the near-IR extinction
(see discussion in Sect. 3.1).

One of the most important  issues directly affected by the correct determination of extinction is the accuracy on distance estimation.
The cluster NGC~3603, for instance,  has distances found in the literature  varying from 6.3 to 10.1 kpc,
a discrepancy mostly due to errors on the reddening correction. 
 Adopting the normal  $R_V$, \cite{2008AJ....135..878M} estimated a distance of 9.1 kpc for NGC 3603,
 while a distance of 7.6 kpc is found if they assume the anomalous extinction law  $R_V$ = 4.3, as proposed by \citet{2000PASJ...52..847P}.
The correct determination of distance is crucial in the estimation of physical parameters of open clusters, like stellar masses and ages. Particularly
in star-forming regions special attention must be paid to the probable occurrence of anomalous or differential extinction. 

Several methods can be employed  to estimate the interstellar extinction in the direction of young star clusters, which can be due to different effects: 
(i) the foreground in a given line-of-sight that is  pervaded by interstellar material in between the cluster and the observer; 
(ii) the presence of a dark cloud associated to the cluster, (iii) the individual extinction caused by circumstellar material.

The use of U-B $\times$ B-V colour-colour  diagram is the classical method for estimating the average extinction towards open 
clusters (e.g. Prisinzano et al. 2011) and it is specially useful on the lack of spectroscopic observations (e.g. Jose et al. 2011).

In addition to colour-colour  diagrams in the UBV bands, two-colour diagrams (TCDs) are also interesting for the study of extinction as
 proposed by Chini \& Wargaw (1990). These diagrams are presented in the form V-$\lambda$ $\times$ B-V, where $\lambda$ refers to different bands.  
The distribution of the stars in TCDs is roughly linear. Anomalies in the extinction law are determined from the comparison between the  distribution 
of field stars (which follows a normal extinction law) with the distribution  of the cluster members (which may follow an anomalous extinction law). 

\cite{2008MNRAS.384.1675J} adopted  TCDs to verify the extinction law in the direction of the cluster Stock~8, for instance. They found the
same  slope for the inner region of the cluster (r $ <$ 6') as well as for a larger area  (r $ <$ 12') that also includes field stars.  
Another example is the cluster NGC 3603 for which \cite{2000PASJ...52..847P} estimated $ R_V $ = 4.3, indicating a remarkable difference 
between the distribution of field stars and members of the cluster on the TCDs.

The aim of the present work is to characterize the extinction in the direction of different young star clusters,
by determining the extinction law  and searching for possible spatial variations. 
We selected four well-known clusters, whose characteristics are compiled in the {\it Handbook
of Star forming regions} edited by Bo Reipurth (2008): 
Berkeley~86 is found in the Cygnus OB1 region \citep{2008hsf1.book...36R};
NGC~2244 is associated to the Rosette Nebula \citep{rz08};
NGC~2264 is related to the Mon OB1 association \citep{2008hsf1.book..966D}; and
NGC~6530 is located near to the Lagoon Nebula \citep{2008hsf2.book..533T}.

The motivation in choosing these well-known objects is to refine the previous extinction determinations, by adopting same 
criteria for  selection and analysis of data sets, in order to compare our results with the characteristics of the clusters environments.
The paper is organized as follows. Section 2  presents the sample and the information available in the literature for the members
of the clusters and the clouds associated to them. Different methods are adopted in Sect. 3, aiming to estimate $R_V$ in the direction of the clusters. 
Section 4 describes an automatic method to fit the ZAMS reddened colours to the observed data, providing
an accurate estimation of visual extinction. In Sect. 5 we develop an analysis of the fractal dimension of the clouds by comparing
the spatial distribution of cluster members with statistical parameters related to clustering. The discussion of the results and the conclusions
are presented in Sect. 6. The colour diagrams utilized for studying the extinction are presented in the Appendix A.


\begin{table}
  \caption{List of the studied clusters.}
{\scriptsize
  \label{tab:table1}
  \begin{center}
    \begin{tabular}{lcccccc}
      \hline
	                &	    l	      &	    b	    &  D   & d &  A$_V$      &F$_{100}$                \\
Cluster          &     ($^{\rm o}$)   &  ($^{\rm o}$)   & ($'$) & (pc) &   mag&   10$^7$Jy/Sr          	    \\ 
\hline 
Berkeley 86  &     76.7     &     +01.3  &   12 & 1585$^a$ &1.7 - 2.5   & 12-15      \\
NGC 2244     &     206.3   &     -02.1  &   29  & 1660$^b$  &  0.3 - 2.2  & 7-20     \\
NGC 2264    &     202.9   &     +02.2  &   39 &   760$^c$   & 0.9 - 3.8    &3-72          \\   
NGC 6530     &      6.1      &     -01.3   &   14 & 1300$^d$ &3.2 - 4.2  & 69-500       \\
  \hline
\end{tabular}
\end{center}
Columns description:  (1) Identification; (2,3) galactic coordinates; 
(4) diameter;  (5) distance obtained from: (a) Bhavya et al. (2007)
(b) Johnson (1962), Park \& Sung (2002);  (c) Sung et al. (1997); 
(d) Mayne \& Taylor (2008). 
}
\end{table}


\section{Description of the Sample}

The list of  clusters and their main characteristics are given in Table 1. This section is dedicated
to summarize the information found in the literature, and to describe the adopted
criteria in selecting the cluster members and  their available observational data.
We also performed an analysis of visual extinction maps and far-IR images of clouds against which the clusters are projected.

\subsection{Selected Clusters}

Berkeley~86 is a particularly small  cluster associated to the Cygnus OB1 region. Its youth was revealed by the presence of O type stars,  discovered by \citet{1974PASP...86...74S}.  Recent results by \cite{bhavya07} indicate a distance of 1585$\pm$160 pc.  They suggest that a low level star formation 
episode occurred 5 Myr ago, and another more vigorous started during the last 1 Myr.
\citet{2008hsf1.book...36R} presents a comparison of  gas and dust distributions in the Cygnus~X region, which is influenced 
by the UV radiation from OB1 association (see their Fig.3). They
show that Berkeley~86 is isolated, projected against an area totally free of gas and dust, suggesting low levels of  extinction for this cluster. 
Yadav and Sagar (2001) argue that Berkeley~86 suffers  non-uniform reddening with colour-excess varying in the range 0.24 $<$ E(B-V) $<$ 1.01,
while Forbes (1981) suggests a more uniform estimation of E(B-V)=0.96$\pm$0.07.

NGC~2244 is located at the centre of the Rosette Nebula. It is a prominent OB association probably responsible for the evacuation of the central part of the nebula. The first photometric study of this cluster \citep{1962ApJ...136.1135J} indicated a colour-excess E(B-V) = 0.46 assuming $R_V$=3.0 that
was later confirmed by \cite{1976ApJ...210...65T} and \cite{1981PASJ...33..149O}. Perez et al. (1987) found anomalous $R_V$ for some of the cluster members and suggested the coexistence of main sequence (MS) and pre-MS stars. Rom\'an-Z\'u\~niga and Lada (2008) suggest ages of 3$\pm$1 Myr.  \cite{wang08} studied the X-ray sources detected by {\it Chandra} in the Rosette region. They verified a double structure
in the stellar radial density profile, suggesting that NGC~2244 is not in dynamical equilibrium.

NGC~2264 is one of the richest clusters in terms of mass range, presenting a  well defined pre-MS \citep{2008hsf1.book..966D}. The estimated
ages vary from 0.1 to 10 Myr (Flaccomio et al. 1997, Rebull et al. 2002).
Surveys in H$\alpha$ and X-rays revealed the presence of about 1000 members. The cluster is observed projected against a molecular cloud complex of \ 
$\sim$ 2$^{\rm o}$ $\times$ 2$^{\rm o}$ and is located at 40' north of the Cone Nebula. The interstellar reddening on this cluster is believed to be low: $A_V$=0.25 estimated by  \cite{1954AJ.....59Q.333W} adopting  E(B-V)= 0.082 and $R_V$=3.08. A more recent work by \cite{2002AJ....123.1528R} derived $A_V$=0.45 from a spectral study in a sample with more than 400 stars. \cite{silvia2010} used the CoRoT satellite to perform synoptical analysis of NGC~2264 with high photometric
accuracy. Their results suggest dynamical star-disc interaction for the cluster members, indicating they are young accreting stars.

In the study by \cite{2008hsf2.book..533T}  of the Lagoon Nebula and its surroundings, the main emphasis is given to NGC~6530, a star cluster 
 associated to an HII region that lies at about 1300$\pm$100 pc. The estimated age of NGC~6530 is less than 3 Myr  (Arias et al. 2007).
Despite being in a line-of-sight that contains a high concentration of gas, NGC~6530 appears decoupled from the molecular cloud once
the cluster members do not seem to be very embedded. This is consistent with evidences that NGC~6530 is inside the cavity of the HII region \citep{mccall90}, indicated by measurements of the expanding gas (Welsh 1983). A range of 0.17 to 0.33 mag has been reported
for the colour-excess in the direction of NGC~6530 (Mayne \& Naylor, 2008). 
\citet{vda97} found a normal extinction law for the intra-cluster
region, while  Arias et al. (2007) suggest $R_V$ = 4.6 for some of the embedded stars. 
Anomalous extinction has been also reported for cluster members individually, for instance the star HD~164740, for   
which Fitzpatrick \& Massa (2009) suggest E(B-V) = 0.86 and 5.2  $<$ $R_V$ $<$ 6.1.


\subsection{Data extraction and Cluster membership}

Two main catalogues of open clusters can be found in the literature: (i) the WEBDA\footnote{
www.univie.ac.at/webda/navigation.html} database and (ii) the catalogue  DAML02\footnote{
www.astro.iag.usp.br/$\sim$wilton} by \cite{2002A&A...389..871D} with the compilations: {\it Tables of membership and mean
proper motions for open clusters},  \cite{2002A&A...388..168D} based on TYCHO2 and \cite{2006A&A...446..949D} based on the
UCAC2 catalogue.

From DAML02  we searched for clusters with ages up to 10 Myr and distances up to 2 kpc, only selecting those which had members with
UBVI photometry and spectral classification available in the WEBDA database. A summary of the characteristics of the  sample selected from these two 
databases is listed in Table \ref{tab:table1}.

In order to search for relevant information  of stars located in the direction of the clusters, three data sets were extracted from WEBDA: (i) UB photometry, (ii) VI photometry, and (iii) equatorial coordinates (J2000). Each source has the same identification in all of these data sets, but the photometry is originated
from different works. For the consistency in the data analysis, we selected only cluster memberships having UBVI photometry
provided by a single work.   The number of studied stars and the corresponding  references to the photometric catalogues are listed in Table 2, which
also gives information on the number of stars having available spectral data.

According to \cite {2001MNRAS.328..370Y}  the correct identification of cluster members is crucial  for the assessment of extinction in that direction and the most reliable selection is based on kinematic studies (proper motion and radial velocity). DAML02 provides a list of stars with JHK photometry -  2MASS catalogue \citep{cutri03} - and also the probability of the star belonging to the cluster (P\%), which is determined from its proper motion. In order to verify if P\% is available 
for  our WEBDA sample, their coordinates were compared with those listed by DAML02. A coincidence of sources was
only accepted for objects having less than 5 arcsec of difference between coordinates.

\begin{figure}
\begin{center}
\includegraphics[width=4cm,angle=0]{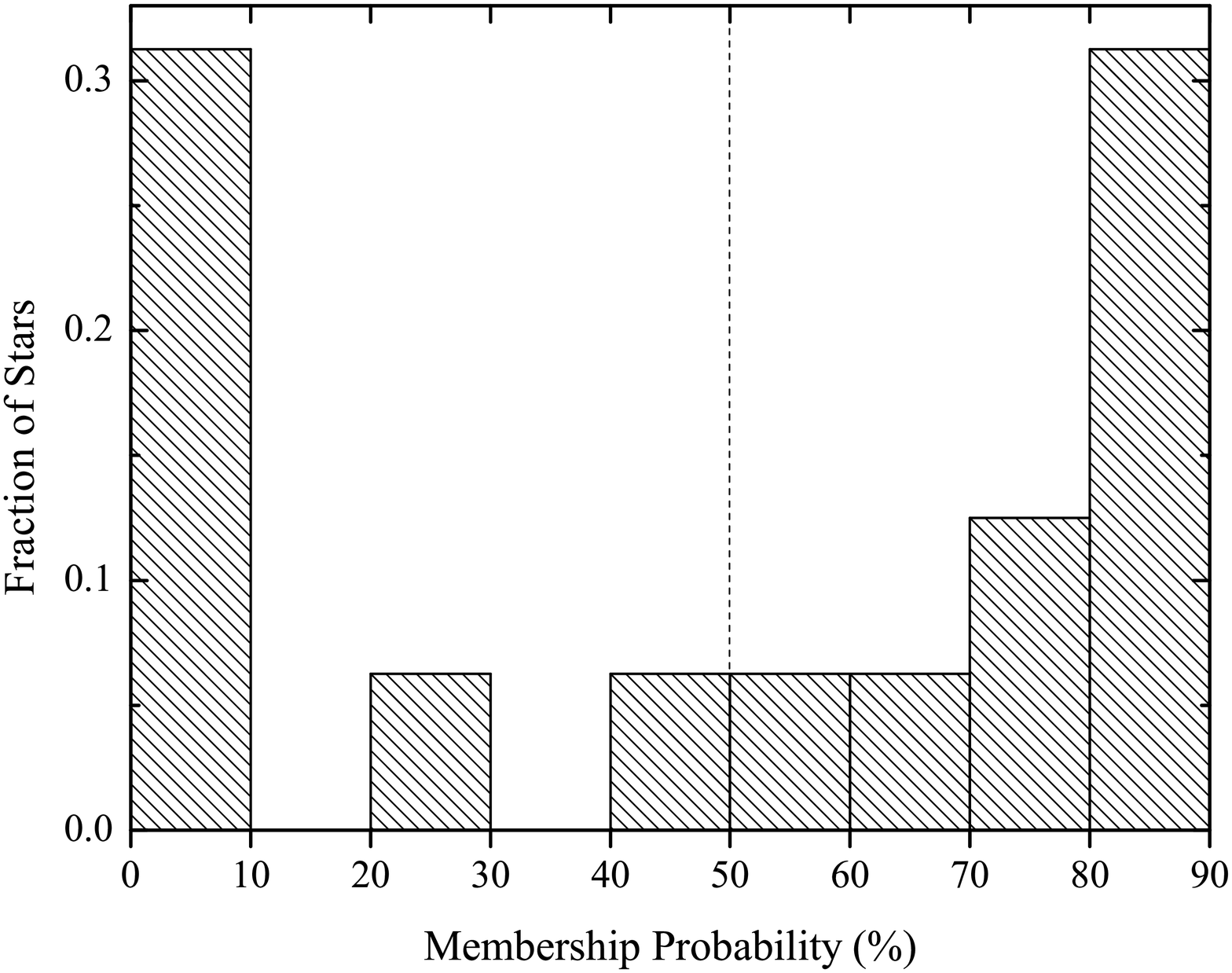}
\hspace{0.5cm}
\includegraphics[width=4cm,angle=0]{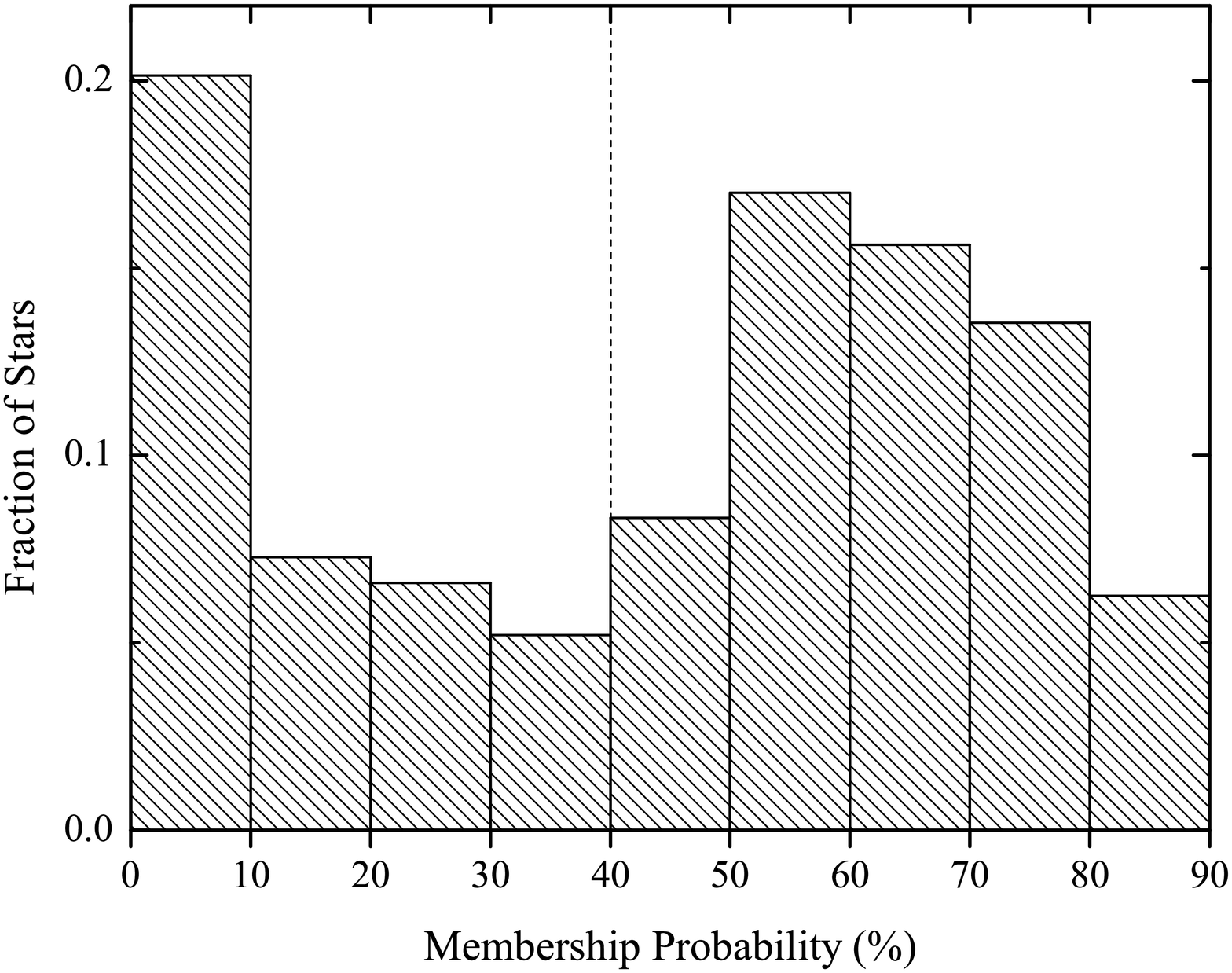}\\
\includegraphics[width=4cm,angle=0]{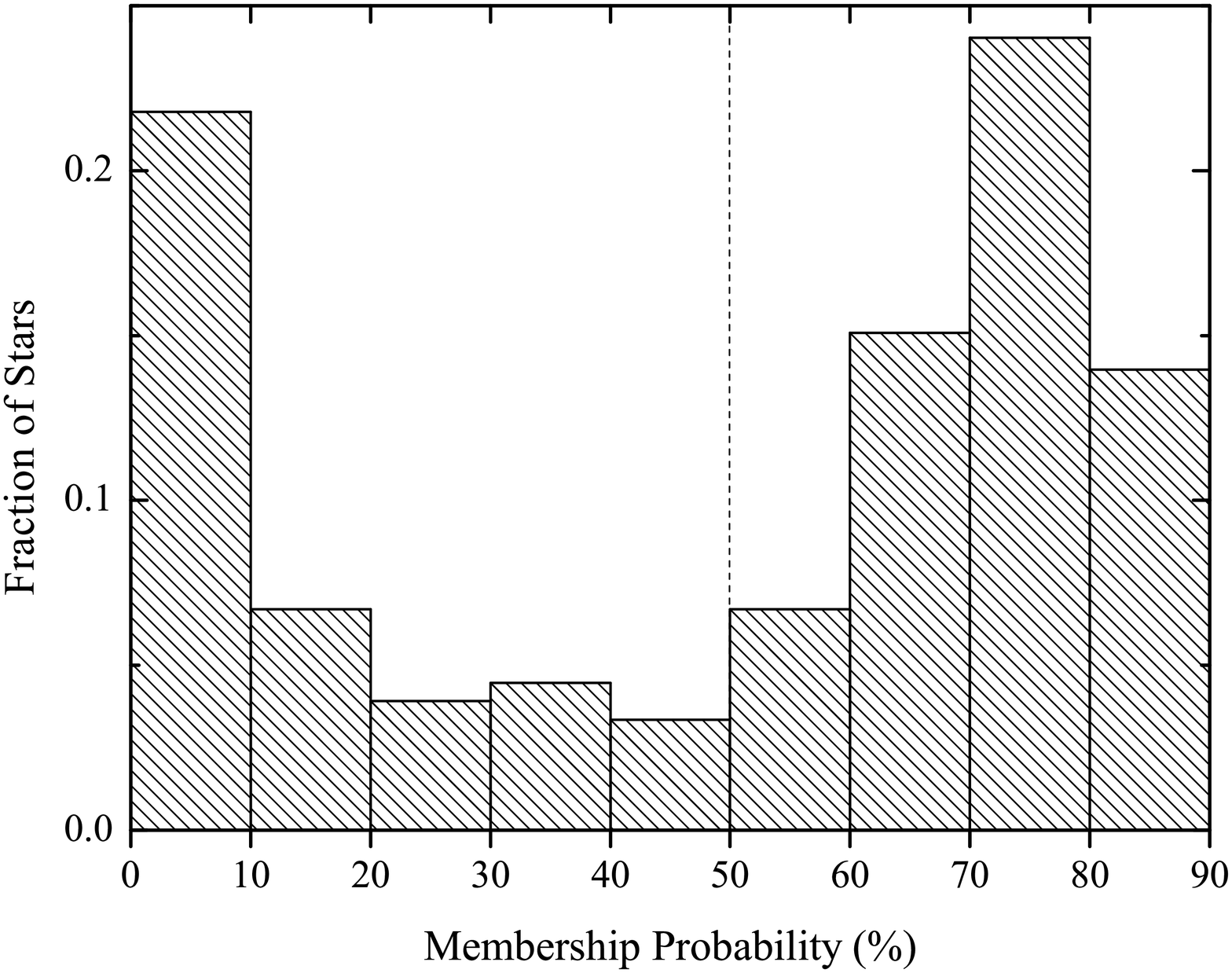}
\hspace{0.5cm}
\includegraphics[width=4cm,angle=0]{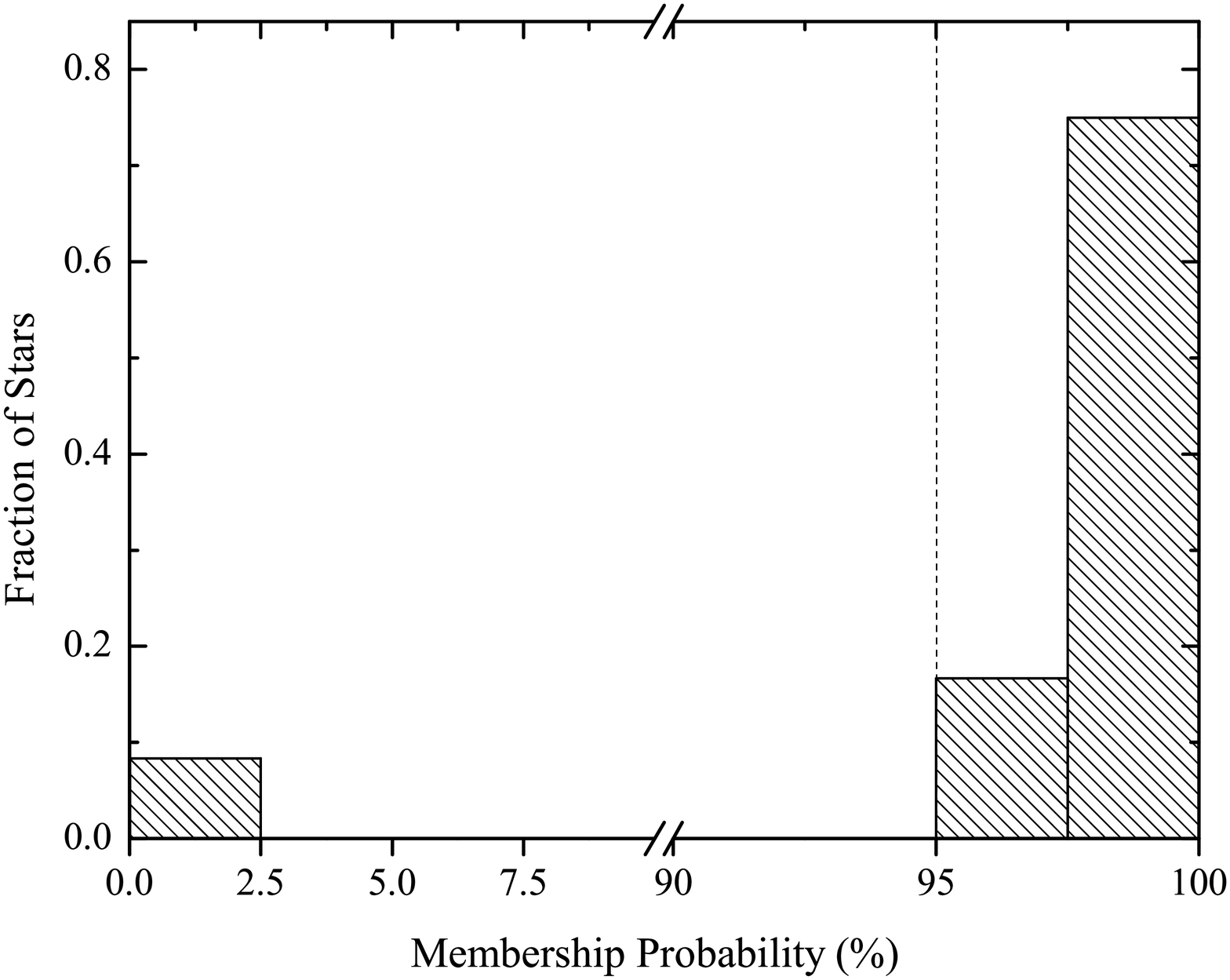}
\caption{ Normalized number of stars distributed according to membership probability (P\%), which is based on their proper motion. 
A dotted line indicates the minimum probability (P$_{min}$) that was adopted to separate the most probable members from field stars.
Top: Berkeley~86 (left) and NGC 2244 (right). Bottom:  NGC~2264 (left) e NGC~6530 (right).}
\end{center}
\end{figure}

The distribution of the number of stars as a function of P\%  was evaluated on basis of  histograms that are presented in Fig. 1.
A bimodal distribution is verified, enabling us  to distinguish between members and possible field stars.
As proposed  by  \cite{2001MNRAS.328..370Y}, the pollution by field stars should be reduced if only stars with high values of 
P\% are considered members of the cluster.
Based on  Fig. 1, for each cluster a minimum probability (P$_{min}$) was adopted to separate field stars, as indicated by dotted
lines in the histograms.

The spatial distribution of  the stars was also checked looking for possible preferential concentrations, as a function of  P\%. However,
no trend was found in the members distribution.

\begin{figure*}
\begin{center}
\hbox{
\includegraphics[bb=28 151 566 688,width=4.3cm,angle=0]{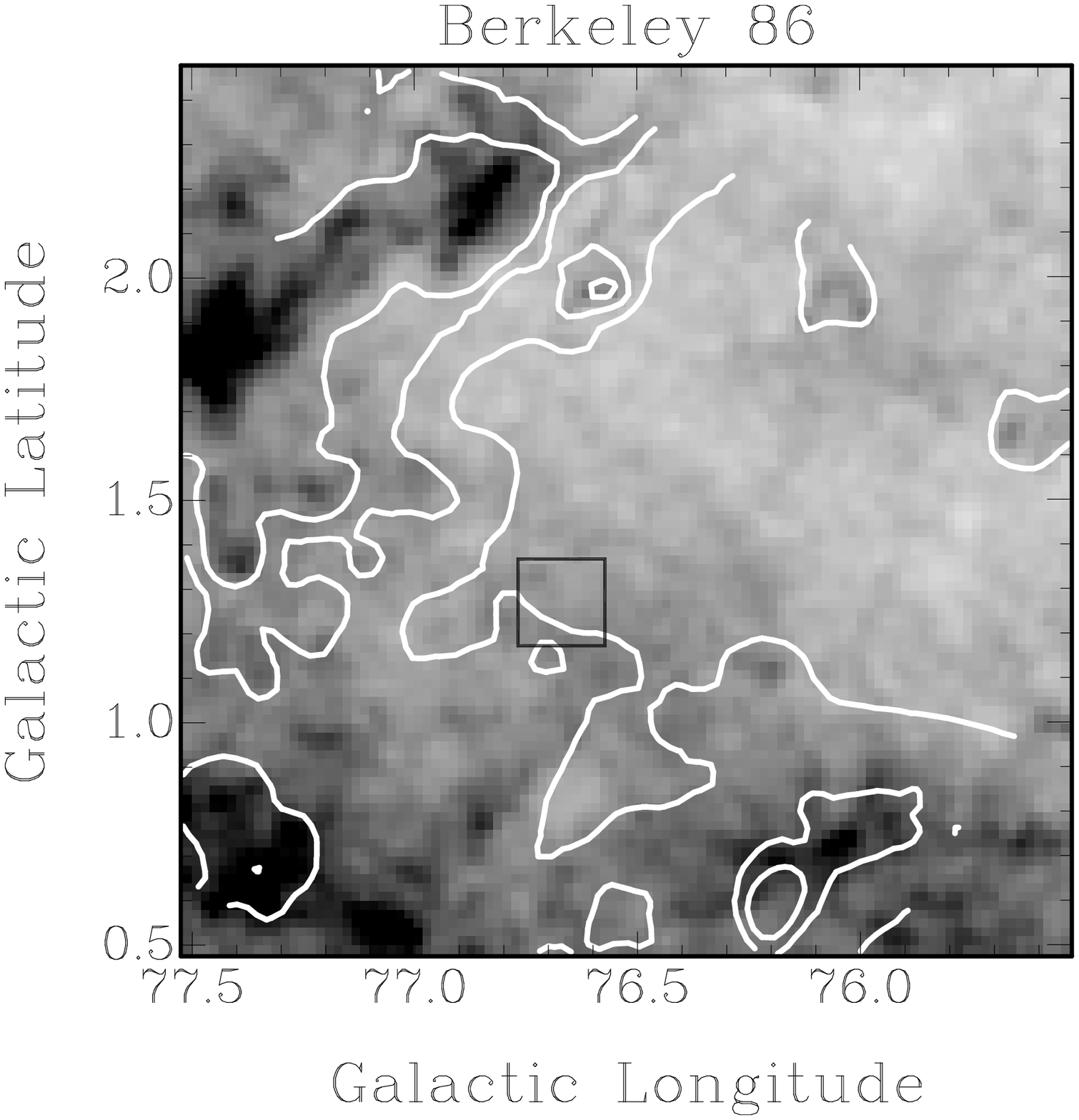}
\includegraphics[bb=28 151 566 688,width=4.5cm,angle=0]{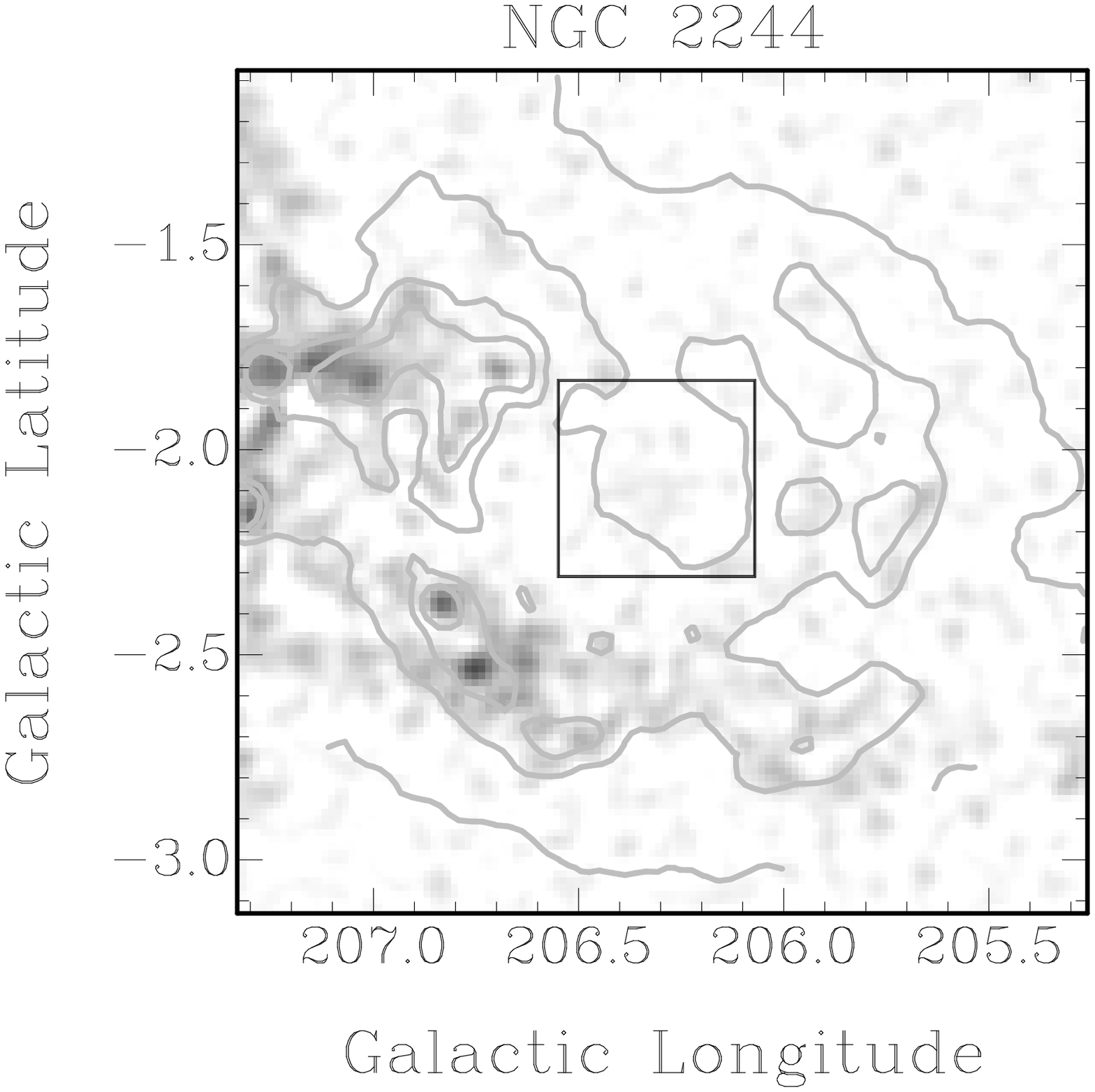}
\includegraphics[bb=28 151 566 688,width=4.5cm,angle=0]{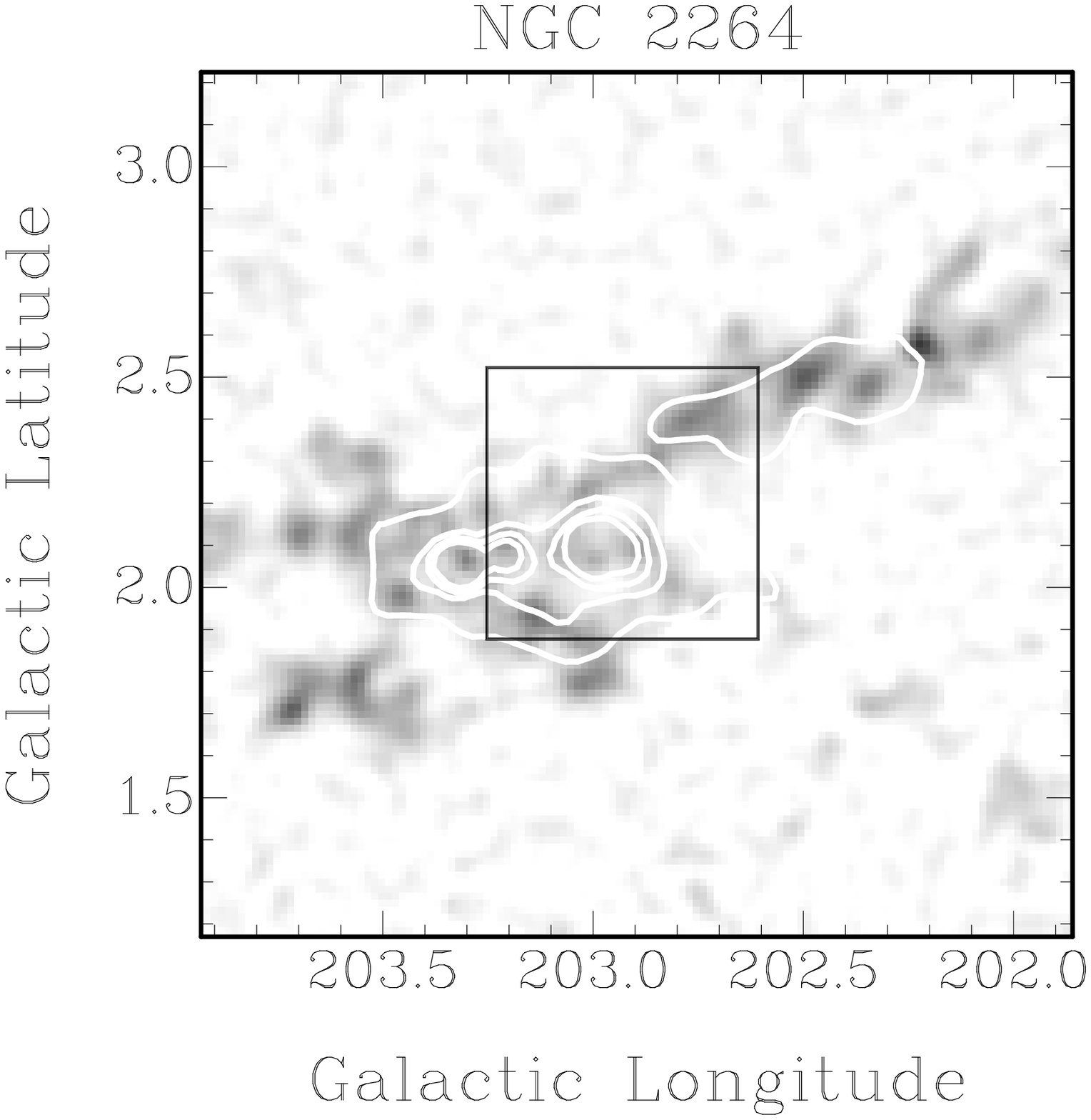} 
\includegraphics[bb=28 151 566 688,width=4.5cm,angle=0]{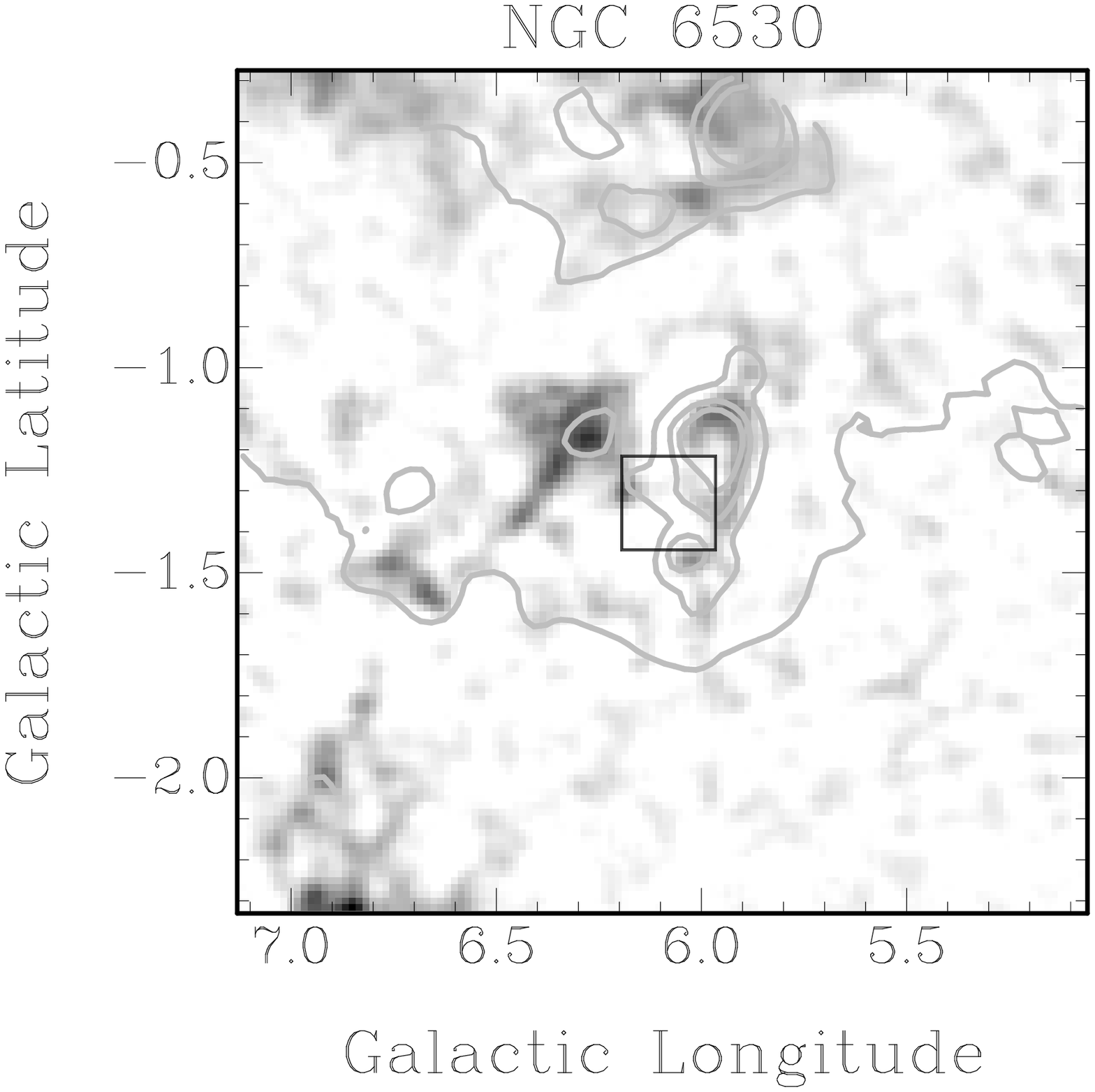} 
}
\caption{Map of visual extinction over-imposed by contours measured from IRIS-IRAS images at 100$\mu$m band. The contours vary from 50 - 350 MJy/Sr 
with steps of 100 MJy/Sr, excepting for  NGC~6530 that has contours starting at 500MJy/Sr with steps of 1000 MJy/Sr. 
The area of the cluster is indicated by the black central square.}
\end{center}
\end{figure*}



\subsection{Comparison with Molecular Clouds}
Extinction caused by interstellar dust found in the line-of-sight can be determined by models that reproduce the  
stars counts in the Galaxy (e.g. Am\^ores \& L\'epine, 2005) or maps of visual extinction (Gregorio-Hetem et al. 1988; 
Schlegel et al. 1998; Dobashi et al. 2005, among others). More recently, colour-excess maps obtained from near-IR
surveys have been used to derive the visual extinction in the direction of molecular clouds, as for instance the Trifid nebula, 
studied by Cambr\`esy et al. (2011) based on {\it 2MASS, UKIDSS}, and {\it Spitzer} data.

Figure 2 displays the position of each cluster projected against  the visual extinction  map obtained from the 2MASS data (K band).
The Catalogue of Dark Clouds by Dobashi et al. (2005) was used to extract the maps for the regions containing our clusters.
In order to compare the extinction with the far-IR flux density of the clouds, the $A_V$ maps are over-imposed by  contours of 
IRAS-IRIS data at 100$\mu$m. A significant variation of both, $A_V$ and far-IR emission, can be found in the direction
of the clusters, excepting Berkeley~86 that appears projected against a uniform field. High levels of far-IR emission are found
in the direction of the clusters NGC~6530 and NGC~2264, which coincide with dense regions in the $A_V$ maps.  
Table 1 gives the ranges of  flux at 100$\mu$m  and  $A_V$  in the direction of the clusters. In spite of the sub-structures
of the clouds being more evident in K band, we adopted the optical (DSS) maps from Dobashi et al. (2005) to measure $A_V$.


\section{Study of the Extinction Law}

\subsection{$R_V$ estimation based on colour-excess}

One of the methods to obtain $R_V$ uses colour-excess, which can be determined  for each cluster member 
that has well-known spectral type.  Intrinsic colours were adopted from 
Bessel et al. (1998), based on surface gravity and effective temperature that were respectively 
extracted from  \cite{1981Ap&SS..80..353S} and  \cite{1987A&A...177..217D} as a function of the spectral type 
and luminosity class available for some stars of the sample.

The expression $\frac{A_{\lambda }}{A_{V}}=a(x)+\frac{b(x)}{R_{V}}$ was adopted from \cite{1989ApJ...345..245C}
in order to obtain relations between $R_V$ and colour-excess. The results are similar to those based on
van de Hulst's theoretical extinction (e.g. Johnson 1968). 
According to \citet{2009ApJ...699.1209F}, these theoretical curves seem to be valid only for  low- to moderate
values of $R_V$, which tend to be underestimated in the case of anomalous high extinction laws. 
They suggested a new formula 
$R_V[K] = 1.36 \frac{E(V-K)}{E(B-V)} - 0.79$, which better reproduces the extinction law in different lines of sight.

Due to the lack of spectral type information for our entire sample, we did not determine $R_V$ for each cluster
member. Instead of that, we  examine the expected ranges of $R_V$  as a function of the distribution
of colour-excesses compared to the theoretical extinction laws on the
E(V-K) $\times$ E(B-V) plot presented in Fig. 3. An illustration of the dispersion found for $R_V$[K]=6 is shown by 
the hatched area, corresponding to the extinction laws given by  \cite{1989ApJ...345..245C} (upper line) and by Fitzpatrick 
\& Massa (2009) (lower line). For $R_V$[K]$<$4 the theoretical lines from both works are quite similar.

Considering the few objects having  known spectral type, this analysis cannot be conclusive for our sample, but gives some 
indication on the variation  of E(B-V) and $R_V$ for each cluster. NGC~2264 shows the largest variation, with E(B-V) ranging from 0.01 to 0.3 mag as  
measured for a sub-sample of 20 stars. 
A degenerate distribution of $R_V$ is found in this case,   spreading from  2.1 to $>$ 6. 
It can be seen in Fig. 3 that NGC~2264 seems to present a bimodal distribution,  
where part of the members follows the normal extinction law, while another part has   R$_V >$ 6.

There are several possible causes to the large variations found in NGC~2264: (i) observational problems that would
give inaccurate optical photometry or spectral type determination; (ii) actual variations in each line-of-sight around the cluster area; 
(iii) anomalies caused by circumstellar environment.  In principle, the two first options may be disregarded, since the observational information
has been checked in other data sets. Furthermore,  there are no correlation between the position of a given cluster member and its $R_V$.
We conclude that a presence of circumstellar matter seems to better explain the anomalies. 
It can also be noted in Fig. 3 that E(V-K)  is particularly large 
for seven objects of NGC~2264, inconsistent with the low E(B-V) values, suggesting  K band excess  that is 
expected in the presence of circumstellar matter. 
Six of these stars have spectral type later than A0, for which it is important
to take into account the errors of spectroscopic data that may cause significant uncertainties on colour excess (see Sect. 3.3).
However, it is interesting to note that some of these objects also show an offset from the normal distribution of the observed
colours on TCDs (see Sect. 3.2 and Fig. A2), which is more consistent with individual anomalies of R$_V$  than possible errors on photometry. 

Smaller variations are found for the other clusters: members of NGC~2244 have $R_V$[K] $\sim$ 3.1; NGC~6530 shows a trend to slightly
larger values (3.1$<$ $R_V$[K]$<$4), while Berkeley~86 tends to have lower values (2.7$< $R$_V$[K]$<$3). However,  
these results are based on few stars  and need to be confirmed by different 
methods of $R_V$  estimation.

\subsection{Two-Colour Diagrams}

In order to search for anomalies in the extinction law, we also constructed TCDs 
following the method  proposed by Chini et al. (1983, 1990) and more recently used by 
\cite{2000PASJ...52..847P}, \cite{2008MNRAS.384.1675J}, Chauhan et al. (2011) and
Eswaraiah et al. (2011), among others. TCDs can be used to perform a qualitative 
 analysis of the nature of the extinction by using the relation:
  $R_{c} = R_{n} \times (a_{c}/a_{n})$, where and $R_{n}$ is the usual normal extinction law,
$a_{c}$ is the slope of the linear fit for the cluster members, and $a_{n}$ is the slope for 
field stars. 

TCDs use V-$\lambda$ colours, where $\lambda$ represents the IJHK bands. In these 
diagrams the intrinsic colours show a linear distribution up to B-V $\sim$ 1.  This limit 
defines a restricted spectral range of validity, where the ZAMS distribution corresponds
 to a straight line. In the TCDs, the distribution of field stars is well reproduced by the 
ZAMS intrinsic colours, once they are reddened by using the mean E(B-V) estimated in 
the direction of the cluster. In this case, the ZAMS fitting is obtained with $R_V$=3.1, 
which we adopted to simulate the distribution of the field stars that must be compared to 
the cluster members.

The fitting of our sample is restricted to the cluster members having P\% $>$ P$_{min}$, 
aiming to avoid confusion with field stars. Since the TCD analysis is only valid for a 
very narrow range of B-V, our calculations are based on objects with spectral type 
earlier than B8. For each studied band, the slope of the linear fit for the 
selected cluster members (a$_c$) was compared to the slope of the normal ZAMS distribution 
(a$_n$). 

Figures A1 and A2 show  the diagrams V-I, V-J, V-H, V-K $\times$ B-V, for which a$_n$ 
is 1.09$\pm$0.02, 1.87$\pm$0.03, 2.42$\pm$0.03 and 2.51$\pm$0.04, respectively. The effect 
of reddening combined with anomalous extinction is illustrated in the V-K $\times$ B-V
diagrams, where extinction vectors were plotted to represent $A_V$ = 2 mag for two choices of
$R_V$, as used by Da Rio et al. (2010), for instance.

Table \ref{tab:table2} presents the results from the TCD analysis giving the fitted line 
slope in each band,  which is used to estimate the mean $R_V$ obtained for the early type 
stars. Within the estimated errors, a good agreement is found in all bands, excepting NGC~6530. 
In this case $R_V$ ranges from 4.5 to 6.2.  This is the only cluster of the sample clearly 
showing anomalous extinction law with significant dependence on wavelength. Berkeley~86 and NGC~2264 
seems to show a normal law. As verified in Sect. 3.1 (Fig. 3), NGC~2244 also has  a bimodal 
distribution in the TCDs: part of the objects follows the normal law, while other 
part shows anomalous extinction.

\begin{table*}
  \caption{Extinction laws from literature and results obtained from the TCD analysis.}
\scriptsize{
  \label{tab:table2}
  \begin{center}
    \begin{tabular}[h]{l|c|c|c|cc|cc|cc|cc}
\hline
Cluster &  N$_{tot}$ & N$_{ST}$ & R$_V$ & $a_{c}$[I] & R$_V$[I]  & $a_{c}$[J] & R$_V$[J] & $a_{c}$[H] & R$_V$[H]  & $a_{c}$[K] & R$_V$[K] \\ [+5pt]
\hline
Berkeley 86 & 16$^a$ &  3$^e$ & 3.0$^i$ & 1.20 & 3.39 $\pm$ 0.11 & 1.98 & 3.29 $\pm$ 0.11 & 2.24 & 2.87 $\pm$ 0.10 & 2.41 & 2.98 $\pm$ 0.10 \\
NGC 2244 & 288$^b$ & 19$^f$ & 3.2 - 3.4$^j$ & 1.14 & 3.22 $\pm$ 0.11 & 2.02 & 3.36 $\pm$ 0.13 & 2.49 & 3.19 $\pm$ 0.10 & 2.54 & 3.13 $\pm$ 0.22 \\
NGC 2264 & 179$^c$ & 24$^g$ & 3.1 - 5.2$^k$ & 1.04 & 2.94 $\pm$ 0.20 & 1.91 & 3.17 $\pm$ 0.08 & 2.39 & 3.07 $\pm$ 0.08 & 2.50 & 3.09 $\pm$ 0.11 \\
NGC 6530 & 24$^d$ & 8$^h$ & 3.1 - 5.4$^l$ & 1.59 & 4.49 $\pm$ 0.14 & 3.45 & 5.73 $\pm$ 0.38 & 4.31 & 5.52 $\pm$ 0.19 & 5.01 & 6.19 $\pm$ 0.32 \\
 \hline
\end{tabular}
  \end{center}
  }
Notes: N$_{tot}$ is the total number of studied stars and N$_{ST}$ refers to those having available spectral type, the respective references to photometric 
and spectroscopic catalogues are: (a) Deeg \& Ninkov (1996); (b) Park \& Sung (2002); (c) Sung et al. (1997); (d) Sung et al. (2000); (e) Forbes et al. (1992); 
(f) Johnson (1962), Morgan et al. (1965), Conti \& Leep (1974), Hoag \& Smith (1959), Wolf et al. (2007); (g) Morgan et al. (1965), 
Walker (1956), Young (1978); (h) Hiltner (1965). The extinction law references are: 
(i)   Bhavya et al. (2007); (j) Ogura \& Ishida (1981), P\'erez et al. (1987); 
(k) Walker (1956), P\'erez et al. (1987); (l) Neckel \& Chini (1981), Arias et al. (2006).
\end{table*}

\subsection{Errors estimation}

Since our work is based on photometric and spectroscopic data from literature, we have evaluated the global uncertainties taking  into account 
the different sources of errors.

According to the references that provided the photometric data (see Table \ref{tab:table2}), the error on the magnitudes  and colours 
is  $\sim$0.08 mag in the worst case. Since this value is smaller than the symbol size on the colour-colour diagrams (see Figs. A1 and A2), 
it does not not affect the linear
fitting used to estimate the variations on the extinction law.

Most objects with known spectral types in our sample are  O or B type stars, whose errors in the characterization of spectral types are not 
significant in our calculation of colour-excess. The only exception is NGC~2264 that has some  A, F or G stars for which we adopted 
a maximum uncertainty of two spectral subtypes.

By combining both, spectroscopic and photometric errors, in order to derive colour-excess,
we estimate the maximum deviation (the worst case) of 0.3 mag on E(B-V) and 0.4 mag on E(V-K) for a G8 type star of 
NGC~2264. Considering that we do not use the colour-excess to derive the extinction law, only performing a qualitative evaluation
of the expected range of R$_V$, these errors do not affect the discussion based on colour-excess.

\begin{figure}
\begin{center}
\includegraphics[width=6cm,angle=270]{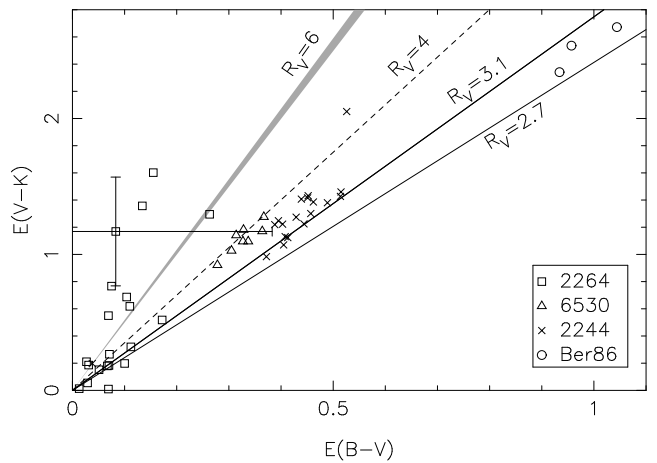}
\caption{Different extinction laws compared to the colour-excess distribution for cluster members that have available 
spectral type. Error bars illustrate the maximum uncertainty expected, if the spectral type 
is not well determined, for a G8 star.
}
\end{center}
\end{figure}


\section{Main Sequence Fitting Algorithm}

The fit of photometric data to the theoretical isochrones is commonly used to determine star clusters parameters. 
In order to avoid subjective results that could be derived from visual inspection, different methods of automatic fitting have been proposed to 
optimize the solution for multi-variable fittings. \cite{mont10}, for instance, developed a technique based on the cross-entropy global optimization
algorithm. This method proved to be a powerful tool to fit the observed data in colour-magnitude diagrams. More recently,
\cite{mont11} used this technique to determine open-cluster parameters by using BVRI photometry, which provides reliable estimation
of distance, extinction, mass and age.

Since we are interested in solving a single variable problem, we decided to adopt a simple method of fitting.  In order to improve the efficiency in determining the visual extinction for the clusters, we developed an algorithm that automatically  fits the theoretical ZAMS curve to the observational data 
 in a colour-colour diagram. The best fitting provides the average $A_V$ in the direction of the clusters. 

We define $Z_{0}[ub_{i}, bv_{i}]$ as the curve formed by the unreddened ZAMS \citep{2000A&A...358..593S} on the  U-B $\times$  B-V
 plan (CC diagram).
The ZAMS is defined by \textit{i}=1..\textit{n}  points (intrinsic colours), while the distribution of observed colours is expressed by $\sigma $, 
where $\sigma _{j}[x_{j}, y_{j}]$ is the position occupied by the star $j$ on the CC diagram, with \textit{j}=1..\textit{m}. In this case, $m$ is
the  total number of  cluster members.

In order to obtain a reddened ZAMS curve  $Z_{A_{V}}[ub, bv]$ by adding the extinction $A_V$  to $Z_{0}[ub_{i}, bv_{i}]$. The extinction was 
calculated by using  the relation $\frac{A_{\lambda }}{A_{V}}$  adopted from \cite{1989ApJ...345..245C} and
the  $R_V$ determined from the TCD analysis (see Sect. 3). The reddened ZAMS was obtained 
by using $R_V$ = 3.1 for NGC~2264, NGC~2244 and Berkeley 86. In the case of NGC~6530 we adopted $R_V$ = 4.5 that was estimated from the V-I $\times$ B-V diagram.

Following the method proposed by Press et al. (1995), a Cubic Spline interpolation is applied to the reddened ZAMS in order to provide a theoretical
 curve to be fitted to the observational data. The best fitting is achieved by searching for an $A_V$ that minimizes the 
distance modulus between the data and the interpolated ZAMS curve, given by:

$$\psi (Z_{A_{V}},\sigma 
)=\sum_{j=1}^{m}{(\bar{ub}_{j}-x_{j})^{2}+(\bar{bv}_{j}-y_{j})^{2}}$$

\noindent where $\lbrack \bar{ub}_{j}, \bar{bv}_{j}\rbrack $ represents the  $Z_{A_{V}}$ point that is closer to $[x_{j}, y_{j}]$.

Figure A3 presents the fitting of the observed colours compared to the curve of intrinsic colours on the U-B $\times$ B-V diagram.
It must be kept in mind that $A_V$  obtained from the ZAMS fitting  is only an average. 
A clear dispersion around this average can be noted in Fig. A3 for  NGC~2264 and NGC~2244. 
According to \cite{1975A&A....43...37B}, a maximum dispersion of $\Delta$E(B-V) = 0.11 may be due to other effects like duplicity, rotation 
and age differences. This expected dispersion is illustrated by two curves displayed in the bottom panels of Fig. A3. 
In this case, only cluster members with P\%$>$P$_{min}$ are plotted.

Clusters with uniform or non-variable reddening present a distribution of observed colours falling into the expected range
of dispersion. The same cannot be said for several members of NGC~2264 and NGC~2244 that have 
colours out of  $\Delta$E(B-V) = 0.11, which suggest  non-uniform reddening.

In summary, the results from the study of extinction law and the main sequence fitting indicate that
Berkeley~86 seems to suffer uniform reddening and normal extinction law. NGC~2264 and NGC~2244 also present normal 
extinction in average, however the reddening is variable for several members of both clusters. 
NGC~6530 shows anomalous extinction (high $R_V$) and uniform reddening, as indicated by the small variation of E(B-V). 

Our results on reddening are in good agreement with those from literature, as can be seen in Table 3 
that gives the colour excesses estimated by: (i) determining the mean value of E(B-V)
on basis of spectral type, as described in Sect. 3.1, and  (ii)  the  ZAMS fitting. 

In order to verify if the colour-excess measured for the stars is compatible with the extinction effects due to the
cloud present in the same line-of-sight, we also present in Table 3 an estimation of  E(B-V) that was converted from 
the extinction map  by adopting R$_V$ = 3.1 and assuming the minimum and 
the maximum A$_V$ values given in Table 1.

Since all objects of our sample are very young, they still are physically associated to the cloud complexes where they have
been formed, but none of them appears to be deeply embedded.
The low extinction of the clouds (A$_V <$ 2.5 mag)  in the direction of Berkeley~86 and NGC~2264
(see Table 1) gives E(B-V) similar to the results found in the literature (respectively Forbes 1981 and Johnson 1962), 
as well as those obtained by us, confirming that they are not surrounded by dense material.
When comparing NGC~2264 and NGC~6530 with their respective clouds, it can be noted
that both are projected against regions showing high levels of dust emission. The flux at 100 $\mu$m reaches
720 MJy/Sr in the case of NGC~2264 and 5000 MJy/Sr for NGC~ 6530, corresponding to high extinction 
(A$_V$ = 3.8 and 4.2 mag, respectively) that suggest the presence of large amounts
of cloud material. However,  both clusters are foreground to these regions,
as indicated by the low level of interstellar reddening previously estimated for NGC~2264 (Walker 1956, P\'erez et al. 1997, 
Sung et al. 1997, Rebull et al. 2002) and NGC~6530 (Mayne \& Taylor 2008) and confirmed in the present work.

Complementing the study of interstellar reddening,  we also performed a fractal
analysis that compares the parameters of the clouds with the spatial distribution of cluster members, which
has been suggested to be a quantitative method  to discuss the relation of 
environmental conditions with the origin of star clusters, as described in the next section.


\begin{table}
  \caption{Colour-excess E(B-V) estimated for the clusters.}
\scriptsize
  \label{tab:table4}
  \begin{center}
    \begin{tabular}[h]{lcccc}
      \hline
               & Berkeley~86 &  NGC~2244 & NGC~2264 & NGC~6530 \\[+5pt] \hline 
Literature & 0.96$\pm$0.07$^a$ & 0.46$^b$ &  0.06 - 0.15$^c$ & 0.17 - 0.33$^d$ \\
S.T. & 0.98$\pm$0.04 &  0.42$\pm$0.06 & 0.13$\pm$0.03 & 0.33$\pm$0.02\\
ZAMS  & 0.89$\pm$0.06 &  0.44$\pm$0.06 & 0.05$\pm$0.06 & 0.21$\pm$0.06 \\ 
$A_V$ Map & 0.55 - 0.81 & 0.12 - 0.71 &  0.28 - 1.22 & 1.03 - 1.35 \\
   \hline
\end{tabular}
  \end{center}
Description of Column 1: {\it Literature}: E(B-V) given by
(a) Forbes (1981),  (b) Johnson (1962),  (c) P\'erez et al. (1987) and Rebull et al. (2002), (d) Mayne \& Taylor (2008);
{\it S.T.}: mean results based on spectral type (see Sect. 3.1); {\it ZAMS}: main sequence fitting (see Sect. 4);
{\it $A_V$ Map}: visual extinction, from Dobashi et al. (2005) map, converted into E(B-V) by adopting normal $R_V$.
\end{table}


\section{Fractal statistic} 

Aiming to investigate a possible correlation between the projected spatial density of the cluster and its corresponding cloud, we 
performed a fractal analysis based on the techniques presented by \cite{hetem93}.  Our results were compared to 
artificial clouds and clusters simulated by \cite{lomax11} (hereafter LWC11). Both 
works discuss the behaviour and evolution of molecular clouds under the interpretation of fractal statistic. 
The fractal dimension measured on realistic simulations of density structures is expected to be related to physical parameters,
which are density dependent like cooling function, 
dissipation of turbulent energy and Jeans limit. However, despite the statistical similarity
between fractal and actual clouds, the link between geometry and physics still relies on empirical concepts.


\subsection{The perimeter-area relation of the clouds}

The clouds studied by LWC11 are artificial, with density profile given by:

$$\rho(r)=\rho_0 \left(\frac{r}{r_0}\right)^{-\alpha}$$
  
\noindent where $\rho_0$ is the density at $r = r_0$, and $\alpha$ defines the density law. 
A possible interpretation of $\alpha$ is related to the definition of fractal dimension
given by \cite{mandel83}: $N{_M}(r)= [\frac{r_0}{r}]^{D_M}$, 
where $N{_M}(r)$ represents the number of self-similar structures observed at scale $r < r_0$.

LWC11 adopted a definition of fractal dimension
similar to the capacity dimension given  by
$D = \frac{log N(r )}{log 1/r}$, where $ N(r)$ is the number of regions of effective side $r$ occupied by data points. A set of points 
has fractal characteristic if the relation  $log N(r)$ $\times$ $log (1/r$) tends to be linear in a given range of $r$. 
The capacity dimension is determined by the slope of this linear distribution \citep{turco97}.

Our statistical analysis of the clouds is based on the visual extinction maps  shown in Fig. 2. The fractal dimension ($D_{2}$)
of contour levels is measured  by using the perimeter-area method described by \cite{hetem93}:
$p \propto a^{D_{2}/2}$ , where $p$ is the perimeter of a given $A_V$ contour 
level and $a$ is the area inside it.

 The perimeter-area dimension depends on the resolution of the maps and their signal to noise relation, 
(S\'anchez et al. 2005). We studied 2$^{\rm o} \times$ 2$^{\rm o}$ regions covered by  
 123$\times$123 pixels, which gives to all the maps the same resolution ($\sim$ 0.02$^{\rm o}$ per pixel).
Since NGC~2264 is less distant than the other clusters, by a factor of about 2, we have degraded the adopted
$A_V$ map in order to simulate a distance of d$\sim$ 1500 pc, which gives a small difference in the measured 
fractal dimension (within the error bar).
In order to minimize errors due to low signal-to-noise ratio (S/N), 
the lower contour level adopted in the calculation of $D_{2}$ was chosen to provide S/N $>$ 10. 
By this way, the low density regions of the maps were avoided in the calculations.


\subsection{The $\mathcal{Q}$ parameter of the clusters}

Since stars are formed from dense cores inside molecular clouds, the spatial distribution of young cluster members
is expected to be correlated to the distribution of clumps.This correlation can be inferred from 
the fractal dimension measured on the cloud compared to the $\mathcal{Q}$ parameter measured on the cluster.
In the technique proposed by Cartwright \& Whitworth (2004), $\mathcal{Q}$ is related to the 
geometric structure of  points distribution
and statistically quantifies fractal sub-structures. Studies on the hierarchical structure in young clusters have used
the $\mathcal{Q}$ parameter to distinguish fragmented from smooth distributions (e.g. Elmegreen 2010).

Two parameters are involved in the $\mathcal{Q}$ estimation:
$\overline{m}$, the mean edge length, which is related to the surface density of the points distribution, and $\overline{s}$ 
that is the mean separation of the points. 
Distributions with large-scale radial clustering, which causes more variation on  $\overline{s}$ than 
$\overline{m}$, are expected to have $\mathcal{Q}$ $>$ 0.8. 
On the other hand,  $\mathcal{Q}$ $<$ 0.8 indicates small-scale fractal sub-clustering, where
the variation of $\overline{m}$ is more important than $\overline{s}$.

The dimensionless measure $\mathcal{Q}$ is given by 
$\mathcal{Q}=\overline{m}/\overline{s}$, with 
$$\overline{m}=\frac{1}{(A_{N} N)^{1/2}} \sum_{i=1}^{N-1}{m_i}$$

\noindent and $$\overline{s} = \frac{2}{N(N-1)R_N}  \sum_{i=1}^{N-1}{}  \sum_{j=1+1}^{N} \vert \overrightarrow{r}_i – \overrightarrow{r}_j \vert$$

\noindent where $N$ is the number of points in the set, $m_i$ is the edge length  of the minimum spanning tree, and $r_i$ is the 
position of point $i$. The area $A_N$ corresponds to the smallest circle encompassing all points, with radius defined by 
$R_N = (\frac{A_{N}}{\pi})^{\frac{1}{2}}$.
To construct the
 minimum spanning tree, we used the algorithm given by \cite{kruskal56}  and to determine the smallest circle encompassing all points we adopted the
 method proposed by \cite{megiddo83}. This technique was adopted by us and employed to the spatial distribution of the cluster members.
 
 The uncertainties in the estimation of $\overline{m}$, $\overline{s}$ and $\mathcal{Q}$ were calculated by using the bootstrapping 
method presented by Press et al. (1995). 
This technique uses the actual set of positions of the cluster members - {$\mathcal{S}_0$} with {\it N} data points - to generate a number {\it M} of synthetic data sets - 
{\it $\mathcal{S}_1$, $\mathcal{S}_2$...$\mathcal{S}_M$} - 
also having {\it N} data points. A fraction {\it f} = 1/e  $\sim$ 37\% of the original points is replaced by random points within the limits 
of the original cluster. For each new set, the parameters $\overline{m}$, $\overline{s}$  and $\mathcal{Q}$ were calculated by adopting 
{\it M}=200.
From these measurements, we derived 1 $\sigma$ deviations
$\Delta\overline{m}$, $\Delta\overline{s}$ and $\Delta\mathcal{Q}$ that are presented in Table 4.

\begin{table}
  \caption{Statistical parameters obtained from fractal analysis}
\scriptsize
  \begin{center}
    \begin{tabular}[h]{lcccc}
\hline
Cluster	& $\mathcal{Q}$ & $\overline{m}$&  $\overline{s}$ & $D_{2}$ \\
\hline
Berkeley~86 &0.73$\pm$0.12 &	0.69$\pm$0.12 &	0.93$\pm$0.14 &	1.39 $\pm$	0.04 \\
NGC~2244 &	0.75$\pm$0.02 &	0.59$\pm$0.02 &	0.78$\pm$0.02 &	1.37 $\pm$	0.05 \\
NGC~2264 &	0.76$\pm$0.02 &	0.61$\pm$0.05 &	0.81$\pm$0.07 &	1.47 $\pm$	0.04 \\
NGC~6530 &	0.85$\pm$0.11 &	0.60$\pm$0.10 &	0.70$\pm$0.11 &	1.34 $\pm$	0.03 \\
\hline
\end{tabular}
  \end{center}
\end{table}

 
\subsection{Comparing clouds and clusters}

Considering that our calculations are made over two dimension maps, while
LWC11 used projections of 3D images, the comparison of measurements of fractal dimension can be done by adopting the equivalence $D_2 \sim D$ - 1. 

Figure 4 shows the $\mathcal{Q}$ parameter as a function of fractal dimension obtained by LWC11.  In order
to illustrate the offset of $\mathcal{Q}$ due to differences on
resolution (or number of points), the clustering statistics for  fractal distributions  
({\it D} = 2.0 - 3.0) of artificial data is displayed for two data sets: N=1024 points and N=65536 points. For comparison with 
our results, the hatched area in Fig. 4 represents the error bars corresponding to the lower 
resolution data set.

Despite the low number of members studied in our clusters, their distribution are consistent with the LWC11 
set of N=1024 points, excepting  
NGC~6530 that tends to be displaced from the cloud-cluster expected correlation. 

Cartwright \& Whitworth (2004) suggested that clusters showing central concentrated distributions have $\mathcal{Q}$ increasing
from 0.8 to 1.5. On the other hand, clusters with fractal sub-structures have  $\mathcal{Q}$  decreasing from  0.8 to 0.45  related to
fractal dimension decreasing from $D_2$ = 2 to $D_2$  = 0.5 (we are using here the same relation between 2D images and 3D distributions
adopted above).
In fact,  NGC~6530 is the only cluster of our sample showing indication of radial profile distribution of stars ($\mathcal{Q}$ = 0.86), which
is not compatible with the fractal structure of its projected cloud ($D_2$ $\sim$ 1.3).
The other clusters show members distribution with
fractal sub-clustering structure, as indicated by their $\mathcal{Q}$ $<$ 0.8 that is in agreement with the fractal dimension measured in the respective projected clouds. 

Several works have statistically proved the correlation between fractal dimension, estimated from cloud maps, and $\mathcal{Q}$ parameter, 
measured on cluster stars distributions. Camargo et al. (2011), for instance, summarizes the concepts first discussed by Lada \& Lada (2003)
suggesting that the structure of embedded clusters is related to the structure of their original molecular cloud. Fractal structures are observed
in clouds showing multiple peaks in their density profile (Cartwright \& Whitworth 2004, Schmeja et al. 2008, S\'anchez et al. 2010, LWC11).  

Based on the comparison with artificial data, we suggest for our clusters a correlation between the fractal statistics of the cloud and the 
cluster members distribution, excepting for NGC~6530. 
In this case, the measured fractal dimension does not indicate an uniform density distribution in the cloud, which is expected for associated clusters having 
central  concentrated distributions. For this reason, our fractal analysis indicates that  NGC~6530 
had a different scenario of formation, when compared to the other three clusters.
 Our argument for discussing the cluster formation comes from the interpretation of the cloud-cluster relation,
 which is derived from the fractal analysis. The meaning of this relation is the comparison of the physical structure
  of the cluster with its remnant progenitor cloud. By this way, the cloud-cluster relation may tell us about the 
  original gas distribution of the cloud that formed the cluster. It also may give us information on how that 
  particular cloud probably evolved since forming the cluster.

The concentrated distribution of NGC~6530 indicates an original material more concentrated than the fractal structure of the cloud
that remained behind the cluster. Possibly this material was contained
in a massive dense core within the cloud, which structure changed due to the process of cluster formation consuming the core.
This scenario is consistent with  the suggestion that NGC~6530 lies within an HII cavity \citep{mccall90}.

\begin{figure}
\begin{center}
\includegraphics[width=6cm,angle=270]{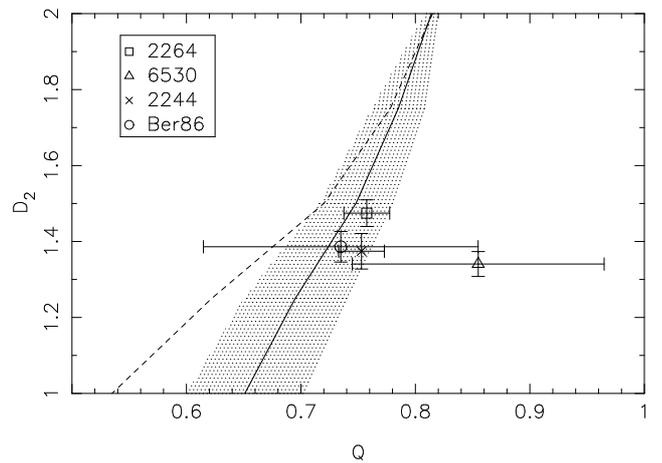}
\caption{Comparison  of our clusters with the $\mathcal{Q}$ parameter  and fractal dimension obtained by LWC11 
for artificial data by using different sets of points: N=65536 (dashed line) and N=1024 (full line and respective deviations shown by the hatched area). 
Symbols with error bars are used
to show our results that were estimated from the spatial distribution of cluster members ($\mathcal{Q}$) and projected clouds ($D_2$).}
\end{center}
\end{figure}

\section{Discussion and Conclusions}

We developed a detailed study of the extinction in the direction of four clusters with ages $<$ 5 Myr, 
located in star-forming regions. The aiming was to search for variable or anomalous extinction that possibly occurs 
for objects associated to dark clouds. In order to verify the characteristics of the clouds that coincide with the projected position of
the cluster, we inspected the visual extinction maps and compared them to the far-IR emission detected in the direction of our sample.

Different methods were used in order to improve previous results on the extinction law $R_V$, providing means to
identify the cause of anomalous extinction. In this comparative study we used the same photometric data base, 
ensuring similar observational conditions for the entire sample. Theoretical expressions for the extinction law were adopted
to evaluate $R_V$ based on the colour-excess E(B-V), which could be determined for some of the cluster members that have 
spectral type available in the literature. Variable extinction law was clearly verified for  NGC~2264, which presented a bimodal 
distribution. Part of the cluster suffers normal extinction and another part shows large dispersion of $R_V$ values, as illustrated
in Fig. 3. 
However, these results are not conclusive since few objects had spectral type well determined, particularly 
Berkeley~86.

A more efficient analysis to estimate $R_V$ and its dependence on wavelength was based on TCDs. The slope of the 
observed colours distribution, compared to the slope of ZAMS colours in Table 2, indicated anomalous extinction
for NGC~6530, which gives $R_V$ = 4.5 measured in the I band, and  $R_V$ = 6.2 in K band.
The other clusters have normal extinction, in average. However, some members of 
NGC~2264 and NGC~2244 appear dispersed in the TCDs, indicating individual anomalous $R_V$, as illustrated in Figs. A1 and A2.
The  TCD analysis confirms the results obtained from the colour-excess analysis, based on spectral type.  
Once there is no relation between the variation of $R_V$ and the spatial distribution of the cluster members, we conclude that 
differences in the  extinction law verified for some cluster members are not caused by environmental differences, 
but are probably due to circumstellar effects. This is particularly in agreement with the results from \cite{silvia2010}
that confirm the accretion activity in circumstellar discs of NGC~2264 members, for instance.

In order to confirm if our results give reliable reddening correction, the $R_V$ estimated from TCDs 
was used to determine the visual extinction that best fits the observed colours of the cluster members.
A ZAMS fitting algorithm  was developed to improve the search for  $A_V$  that minimizes the distance of the colours
distribution to the reddened ZAMS curve, which was reproduced by Spline interpolation. The best fitting provided a mean value for
E(B-V), which was compared to the results from other methods (Table 3).
All the colour-excess estimations are in good agreement, excepting the E(B-V) derived from $A_V$ maps 
mainly for NGC~2264 and NGC~6530. These clusters are not embedded on their respective clouds, which is consistent with
the extinction levels found for the cluster members being lower than that 
measured on the extinction maps for the background stars.

Fractal analysis was performed to investigate the sub-structures of the clouds and to compare them with statistical parameters
of the cluster members distribution. The estimation of the $\mathcal{Q}$ parameter indicated that NGC~6530 has a 
radially clustered spatial distribution, while the other clusters have  fractally sub-clustered distribution of members. 
The fractal dimension ($D_{2}$) measured on the $A_V$ maps, which is related to the cloud geometry, was compared to $\mathcal{Q}$ 
measured on the clusters distributions. Excepting NGC~6530, the clusters have $\mathcal{Q}$ parameter compatible with the fractal 
dimension of the corresponding clouds,
similar to the distribution of artificial data.
We  interpreted this result as a correlation between cloud structure and cluster
members distribution, which is similar for Berkeley~86, NGC~2244 and NGC~2264. 
On the other hand, NGC~6530 does not show this cloud-cluster relation,
indicating that it was formed from a gas distribution more centrally concentrated, different
of the fractal sub-structures found in the remaining cloud.

In fact,  even under anomalous $R_V$ = 4.5, NGC~6530 does not suffer high extinction, as inferred 
from the average colour-excess ($A_V \sim$ 1 mag), which is incompatible with the high levels of extinction shown in
the dark clouds map or the far-IR emission maps. We conclude that anomalous extinction in this case is not due to
interstellar dense regions. A tentative explanation is the depletion of small grains due to evaporation
that occurs under radiation from hot stars, which is consistent with the location of NGC~6530 in an HII cavity. Neither the circumstellar
effects can be disregarded, since the grain growth is expected to occur in protoplanetary discs. In this case, our results
are also in agreement with the polarimetric results from McCall et al. (1990) that suggested the anomalous $R_V$ is
due to circumstellar effects.

We also verified that the fractal analysis used to investigate the cloud-cluster relation may give some indication about the 
scenario of the cluster formation.
Once the role of this cloud-cluster relation  is  comparing the physical structure of the 
cluster with its parental cloud, the interpretation of this relation may tell us about the original gas distribution of the 
forming cloud and how the particular cloud may have evolved since the cluster was formed.
It must be kept in mind that suggesting a cloud-cluster connection  based on the correlation of $\mathcal{Q}$  with $D_{2}$ 
is only speculative in the case of our sample, since more points are required to have a robust analysis. However, 
it is interesting to note the observed trend when comparing calculations performed for data sets from 
unrelated origins. 
This trend suggests a connection of stars clustering  with sub-structures
of the clouds, comparable to the artificial clouds and their derived clusters, studied by LWC11.  
For this reason, we conclude that NGC~6530 shows actual differences when compared with the other clusters, which is in
agreement with previous results from literature.

An interesting perspective of this work is to extend our study to a larger number of clusters, mainly those that have not been studied as well as our sample has been.

\begin{acknowledgements}
     Part of this work was supported by CAPES/Cofecub Project 712/2011. BF thanks CNPq Project 142849/2010-3.
This publication makes use of data products from the Two Micron All Sky Survey, which is a joint project of the University of Massachusetts and the Infrared Processing and Analysis Center/California Institute of Technology, funded by the National Aeronautics and Space Administration and the National Science Foundation.
\end{acknowledgements}

\begin{appendix} 
\section{Colour Diagrams}
\begin{figure*}[h]
\begin{center}
\includegraphics[width=8.5cm,angle=0]{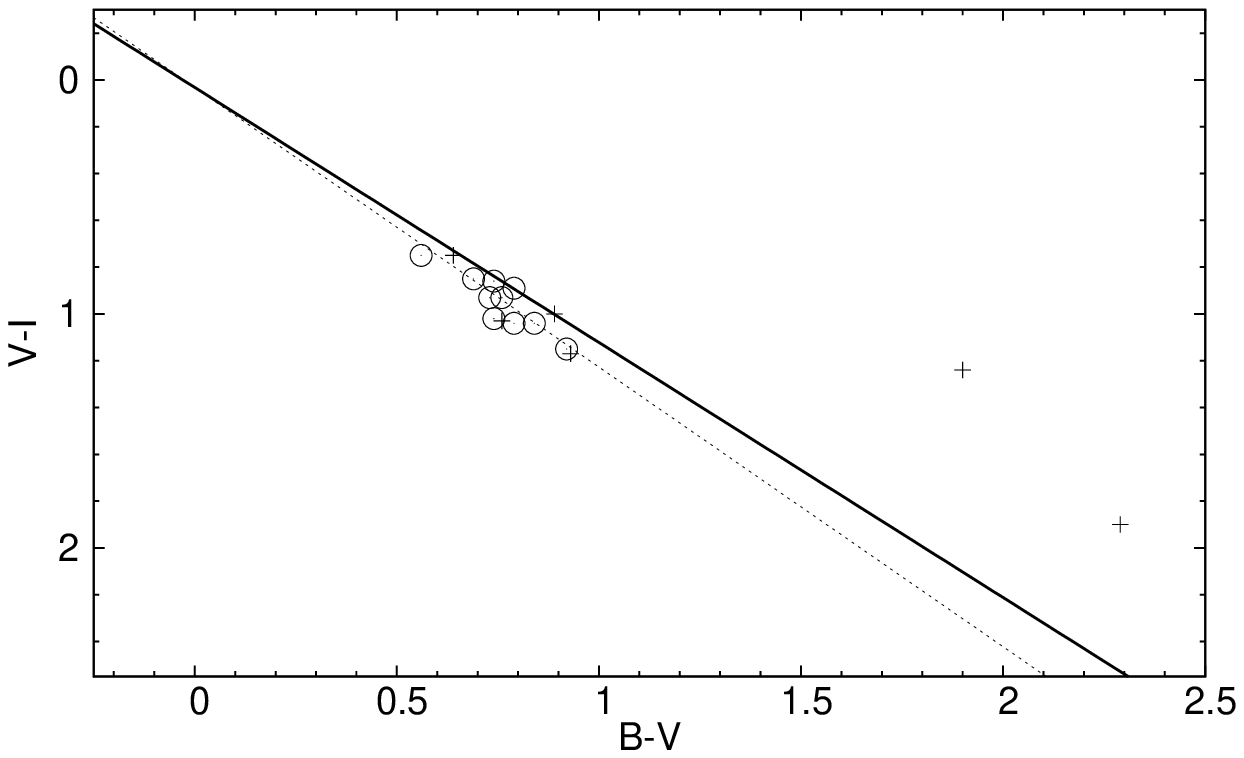}
\includegraphics[width=8.5cm,angle=0]{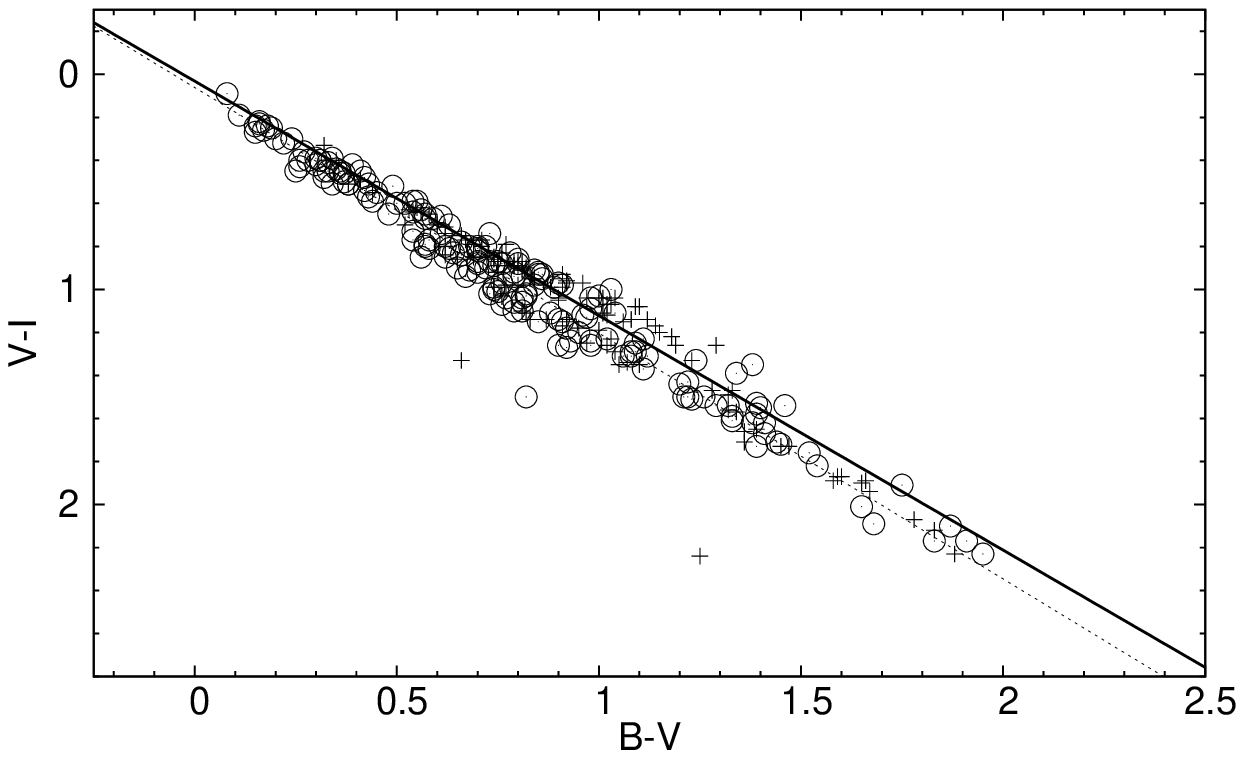}
\includegraphics[width=8.5cm,angle=0]{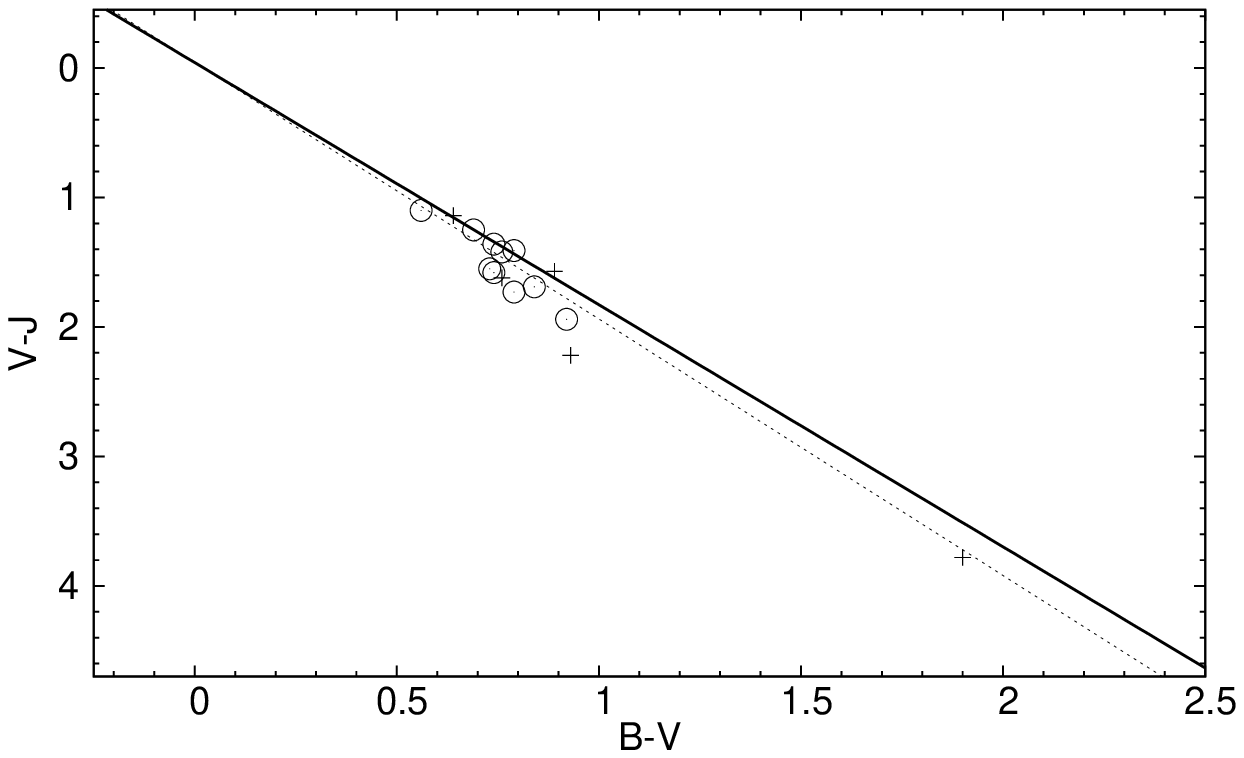}
\includegraphics[width=8.5cm,angle=0]{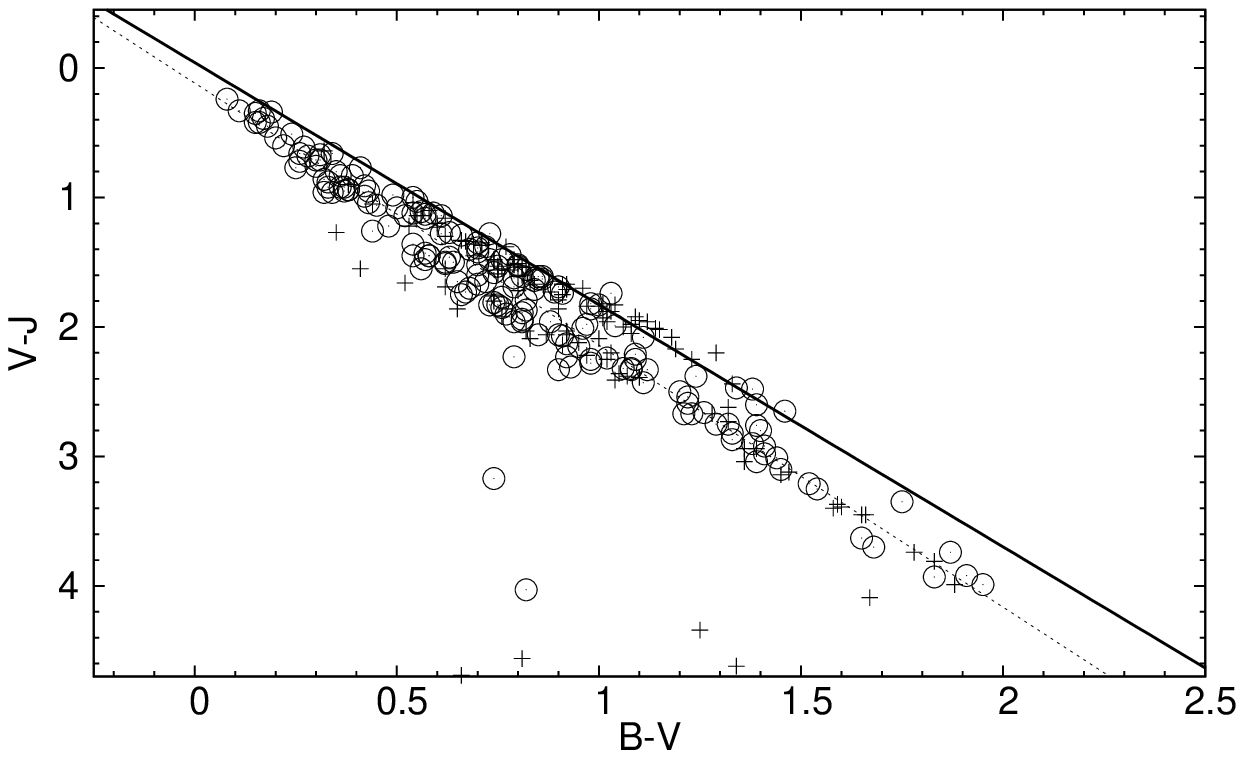}
\includegraphics[width=8.5cm,angle=0]{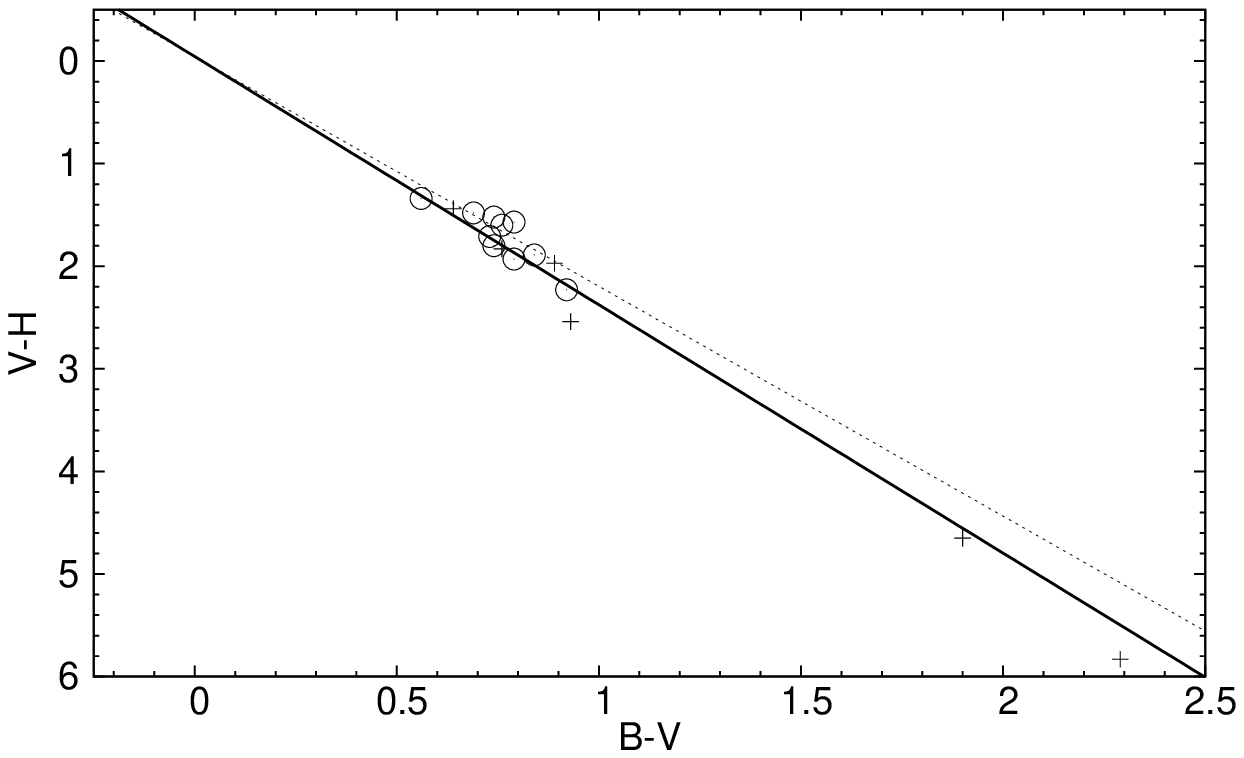}
\includegraphics[width=8.5cm,angle=0]{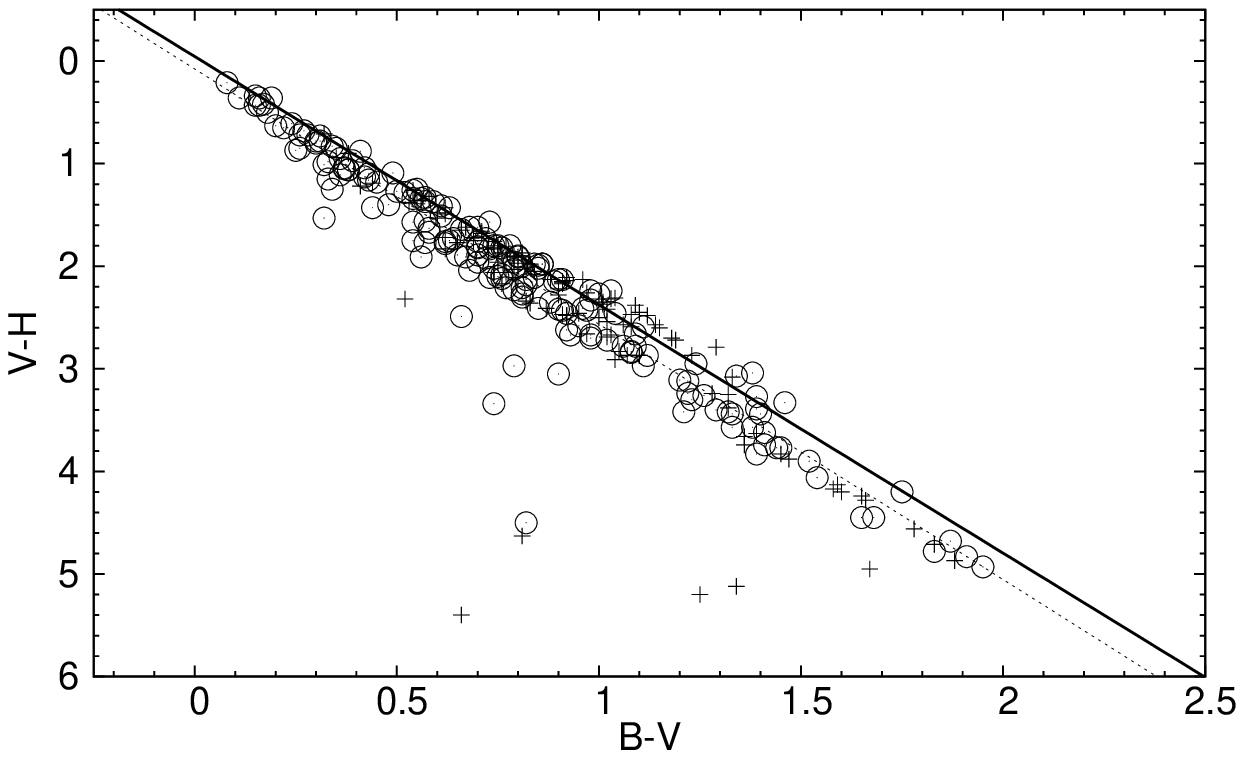}
\includegraphics[width=8.5cm,angle=0]{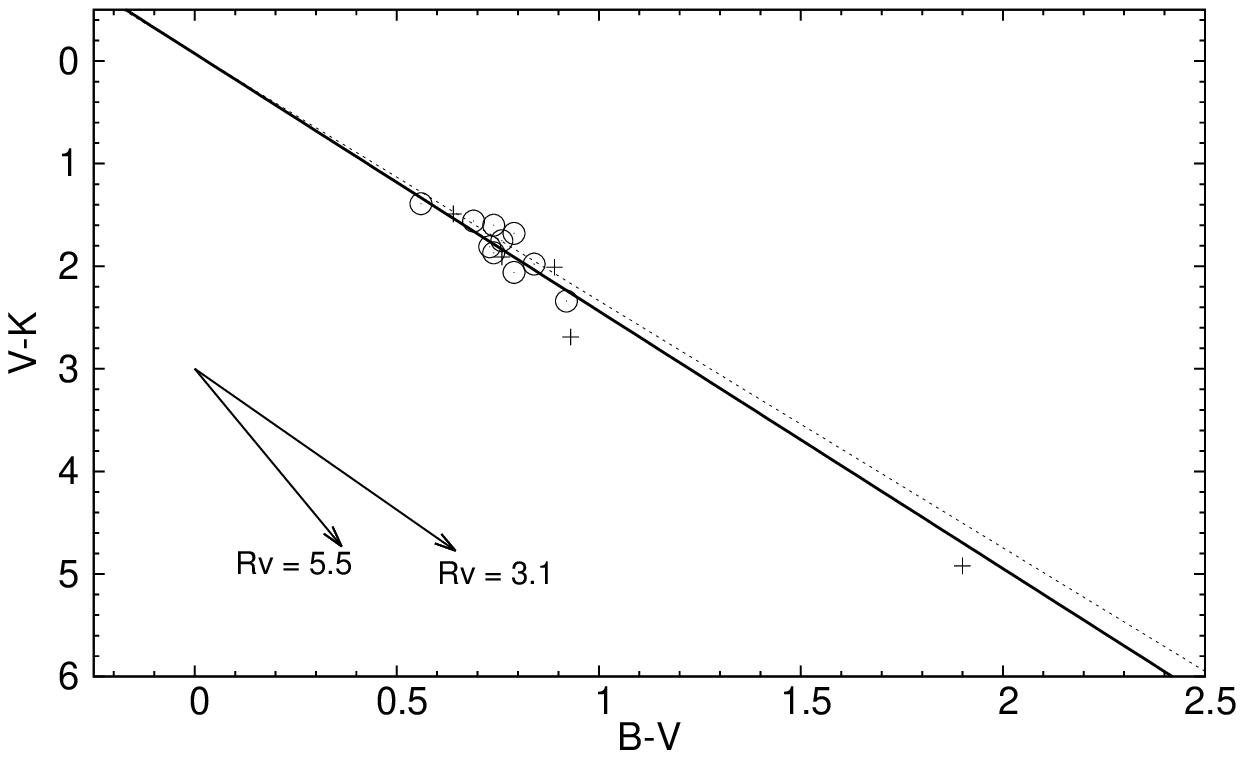}
\includegraphics[width=8.5cm,angle=0]{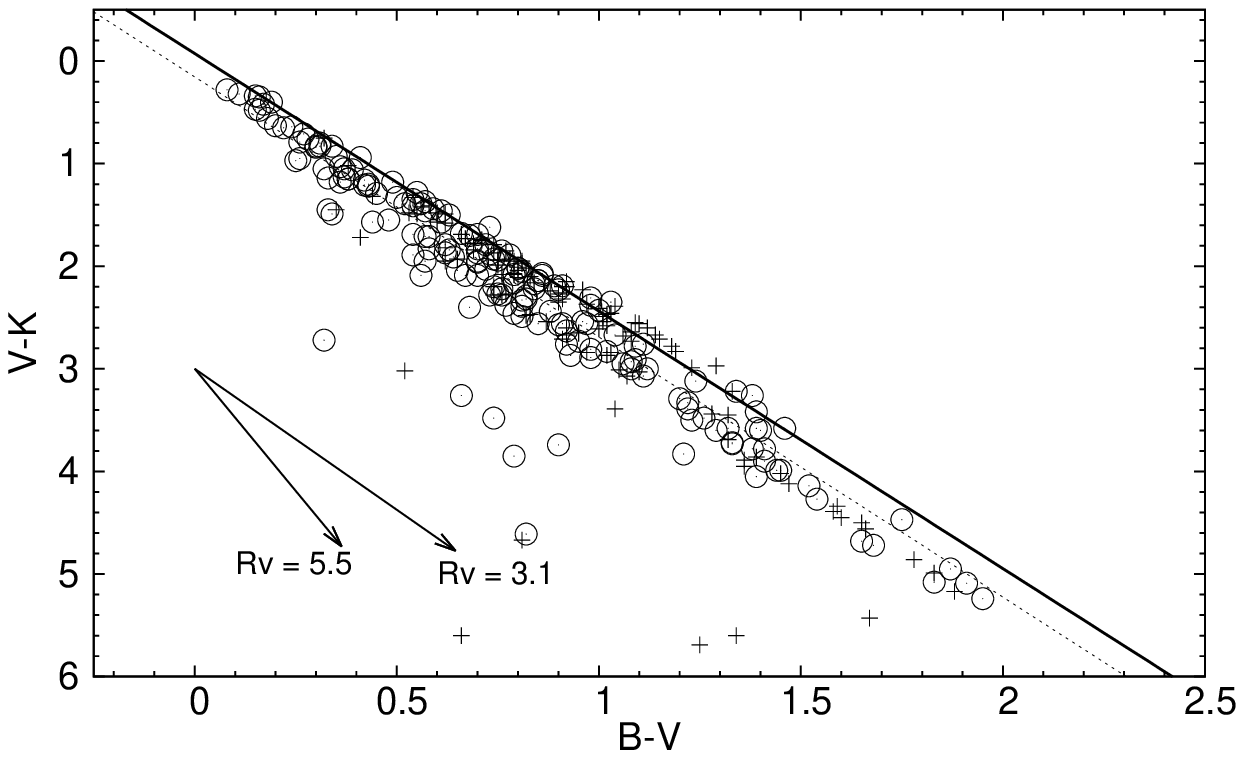}
\caption{TCDs for the clusters Berkeley~86 (left) and NGC~2244 (right). The solid line shows a normal extinction law, while the fitting for the cluster members is shown by a dotted line. Two extinction vectors ($A_V$ = 2 mag) related to different values of $R_V$ are plotted in the bottom panels, for illustration. Open circles indicate the members and crosses show the stars with P\%$<$ P$_{min}$.}
\end{center}
\end{figure*}

\begin{figure*}[ht]
\begin{center}
\includegraphics[width=8.5cm,angle=0]{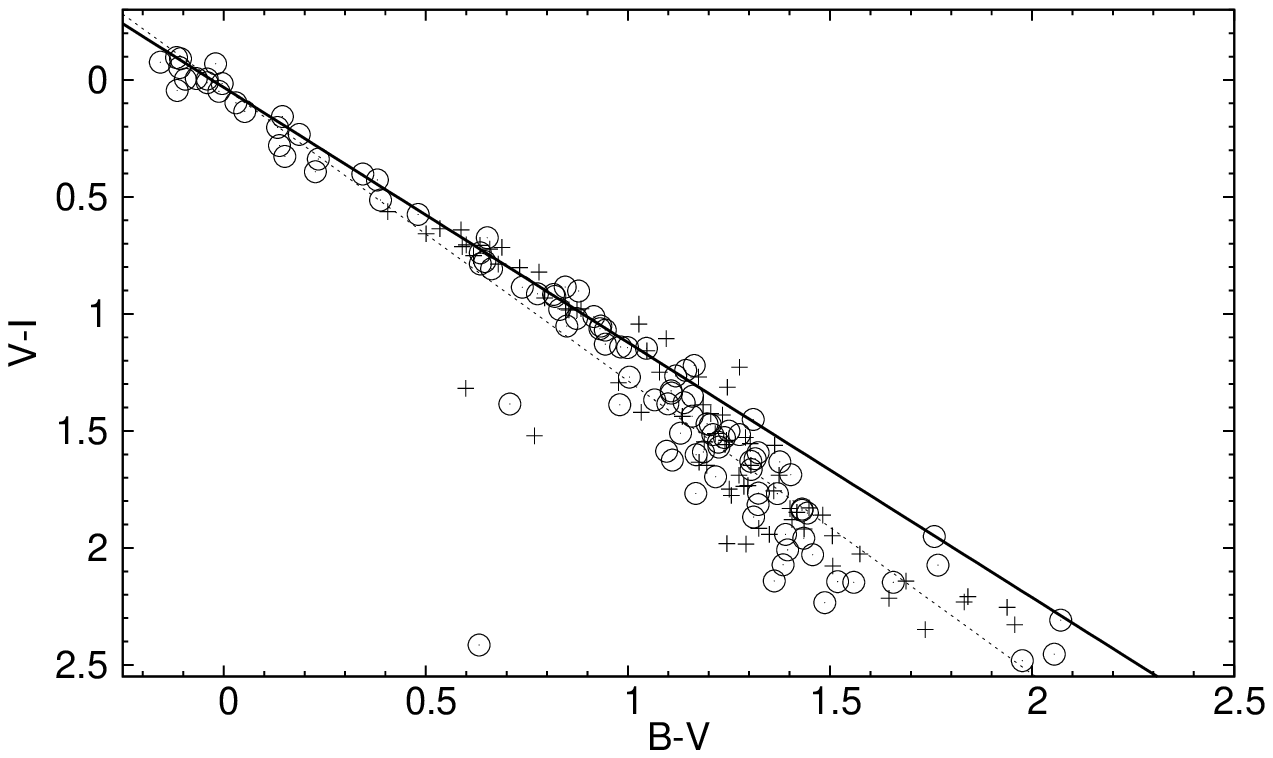}
\includegraphics[width=8.5cm,angle=0]{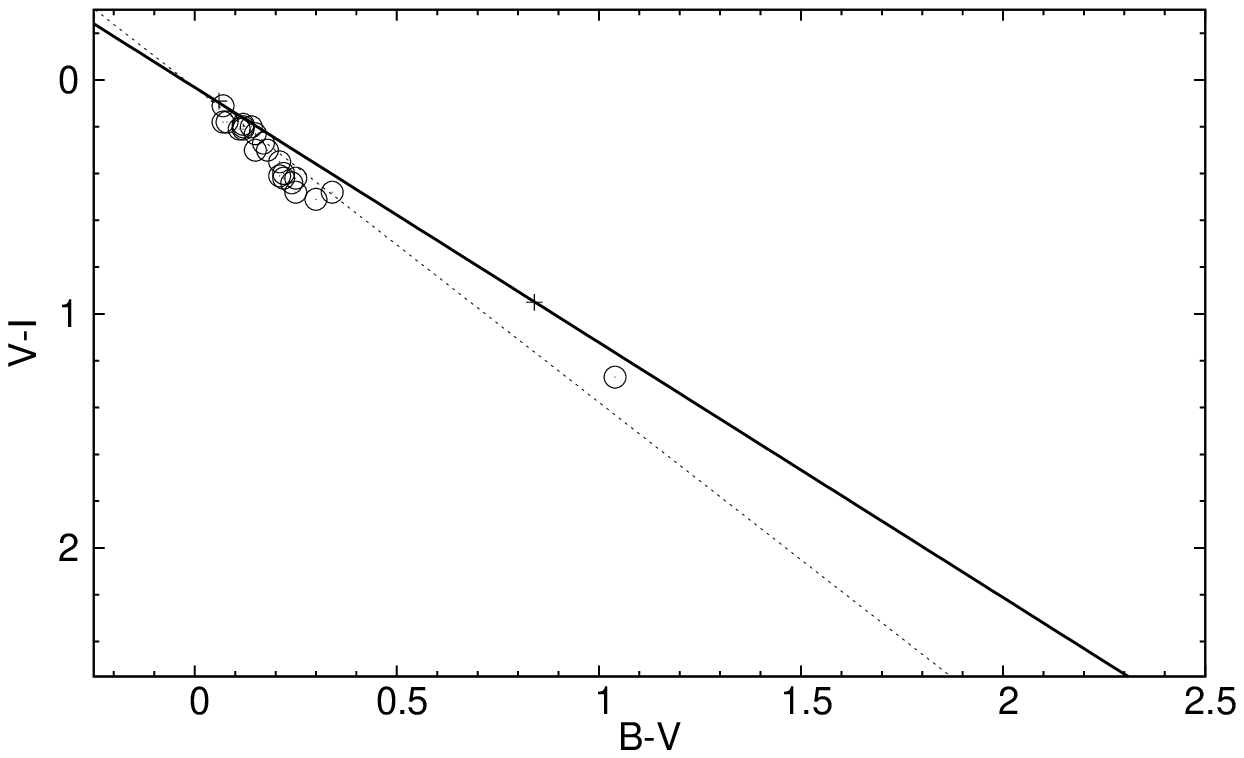}
\includegraphics[width=8.5cm,angle=0]{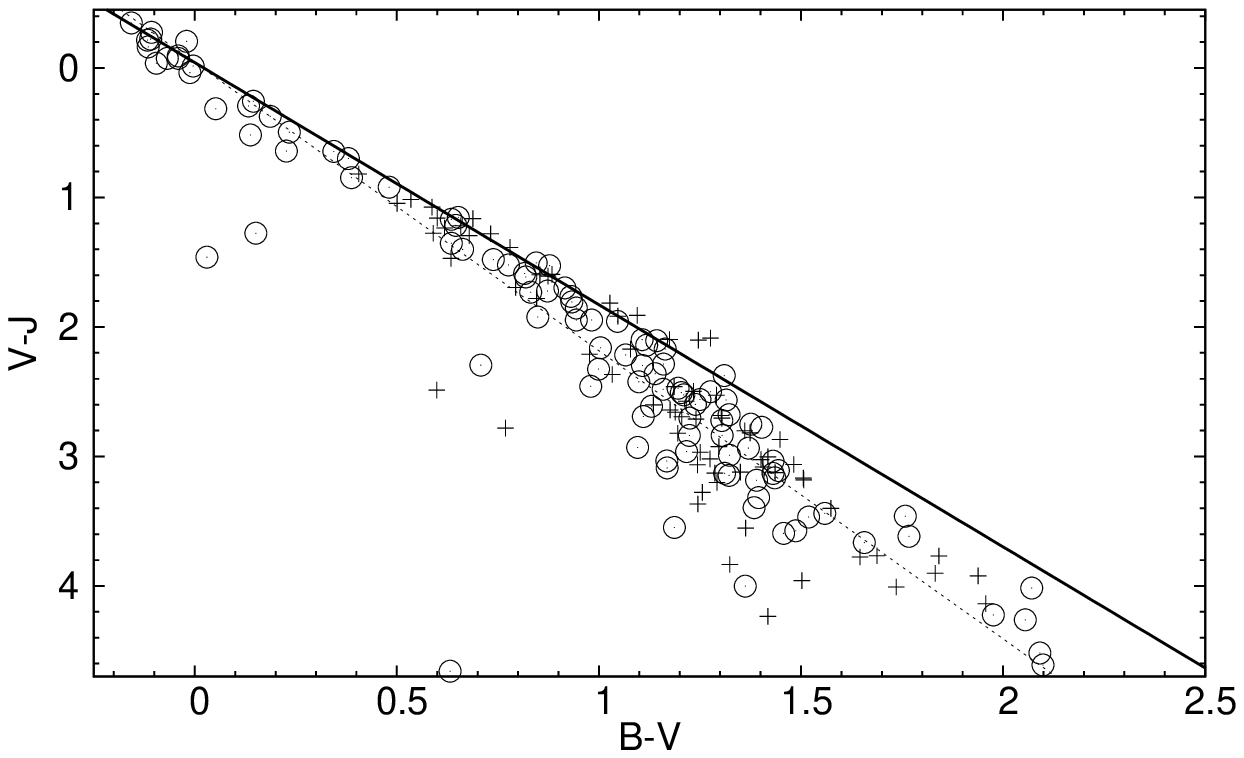}
\includegraphics[width=8.5cm,angle=0]{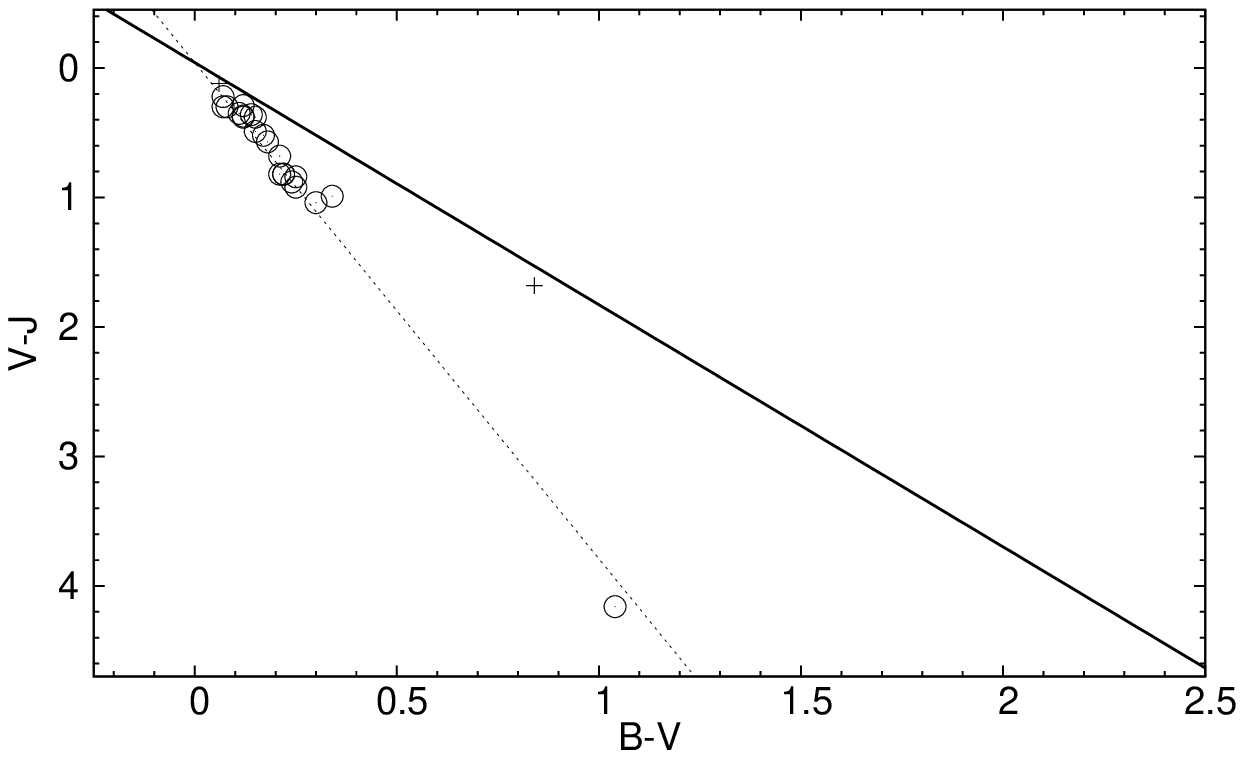}
\includegraphics[width=8.5cm,angle=0]{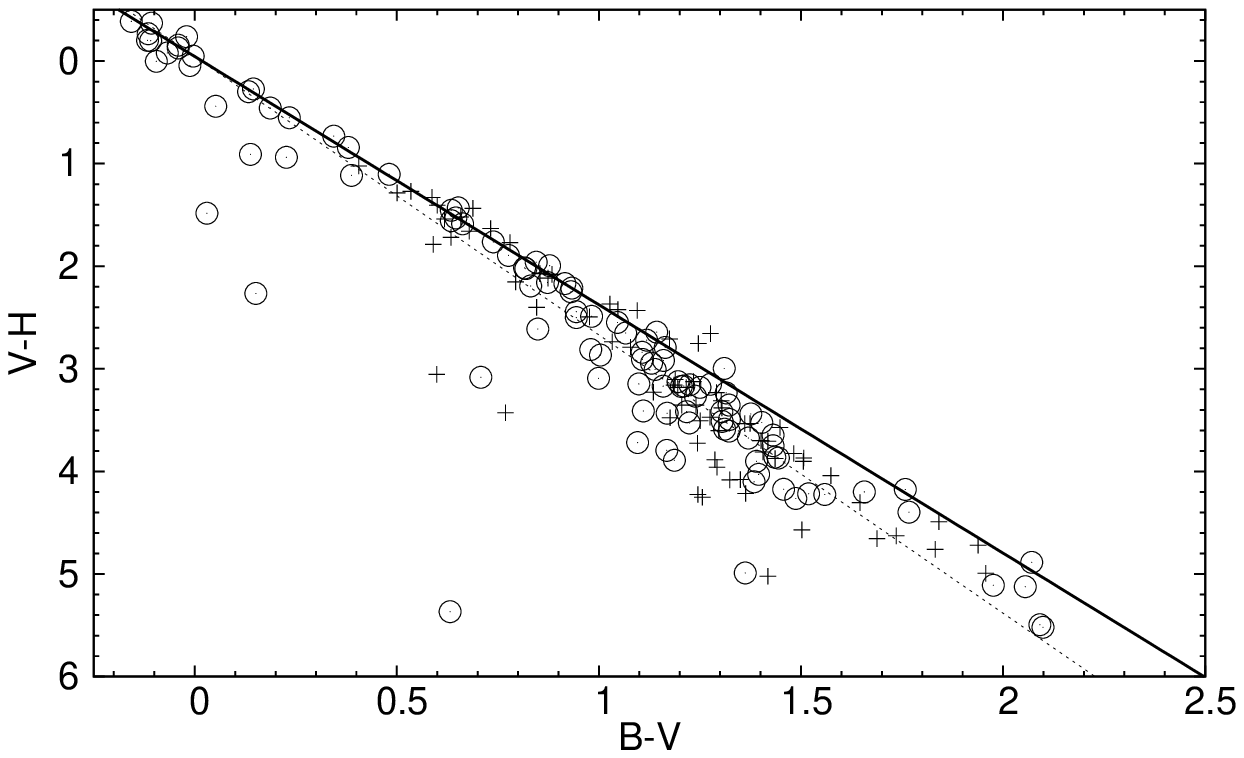}
\includegraphics[width=8.5cm,angle=0]{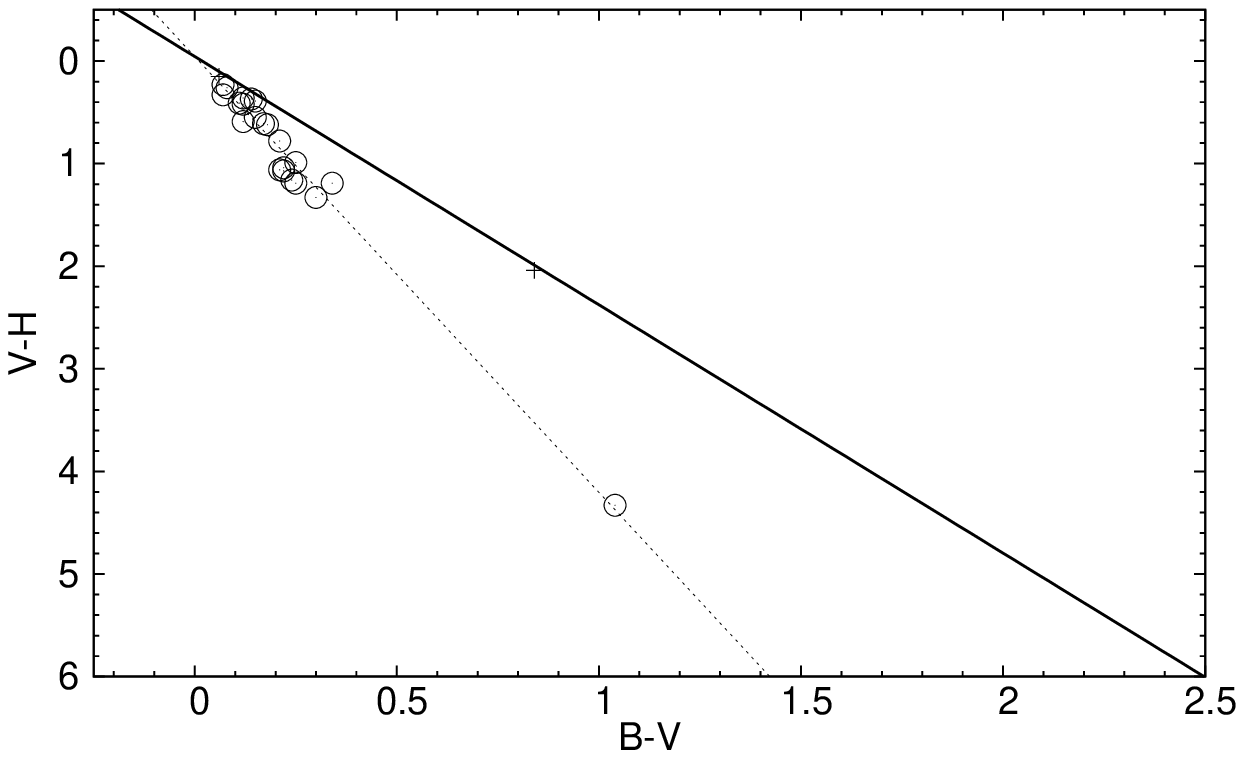}
\includegraphics[width=8.5cm,angle=0]{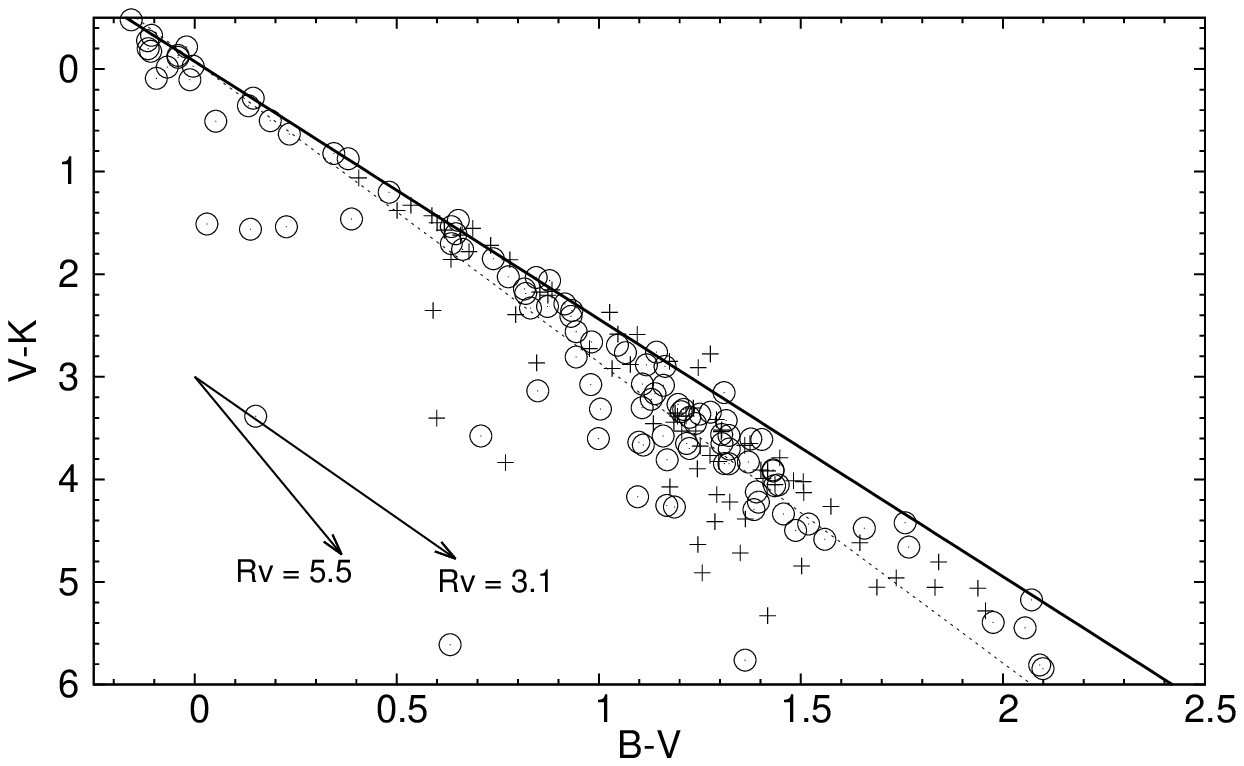}
\includegraphics[width=8.5cm,angle=0]{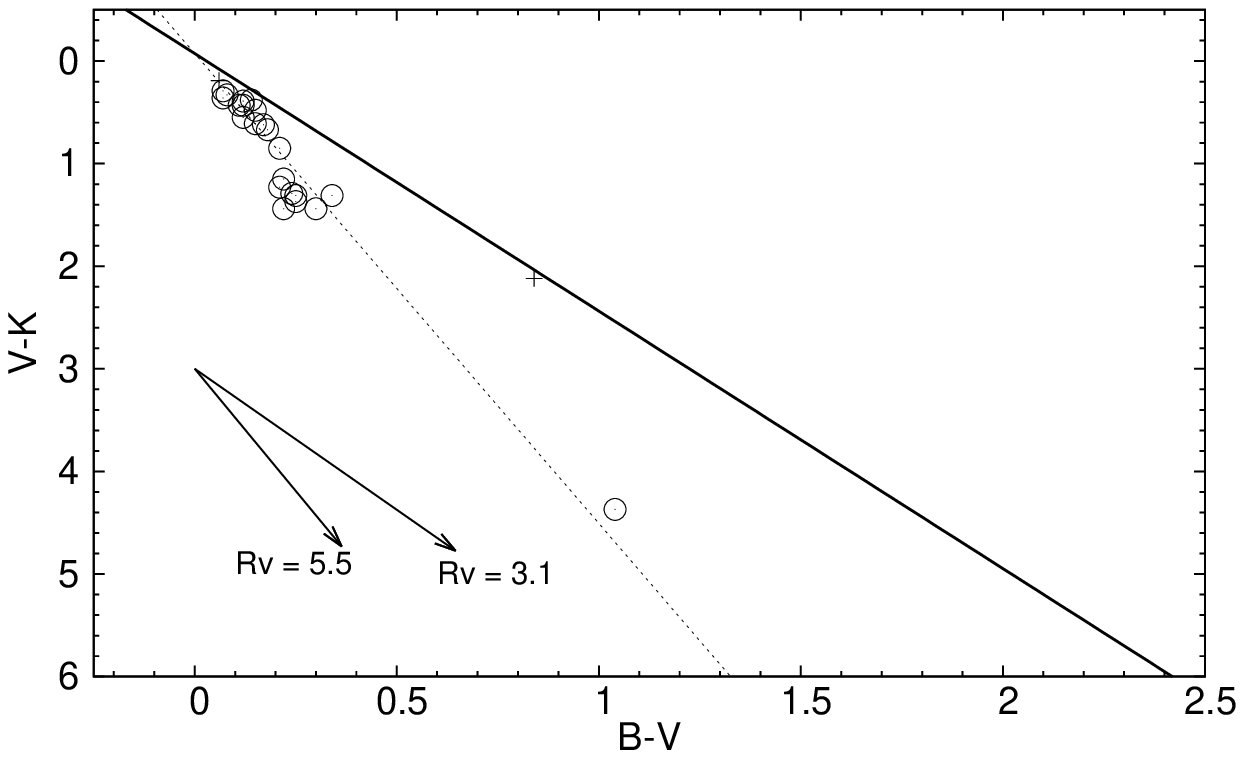}
\caption{TCDs for  NGC~2264 (left) and NGC~6530 (right)}
\end{center}
\end{figure*}
\begin{figure*}
\begin{center}
\includegraphics[width=4.5cm,angle=0]{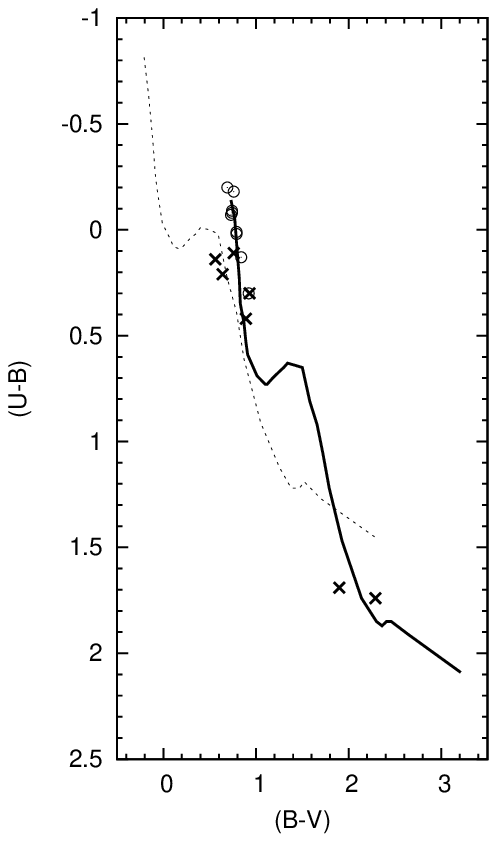}
\includegraphics[width=4.5cm,angle=0]{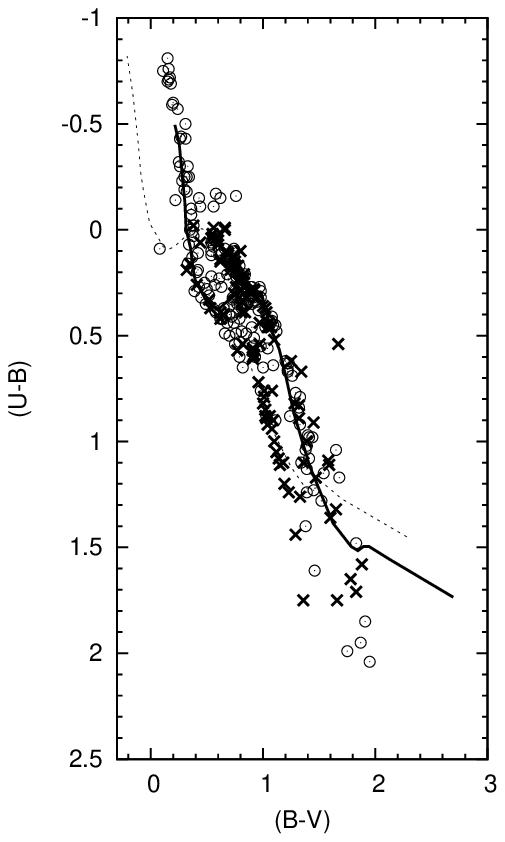}
\includegraphics[width=4.5cm,angle=0]{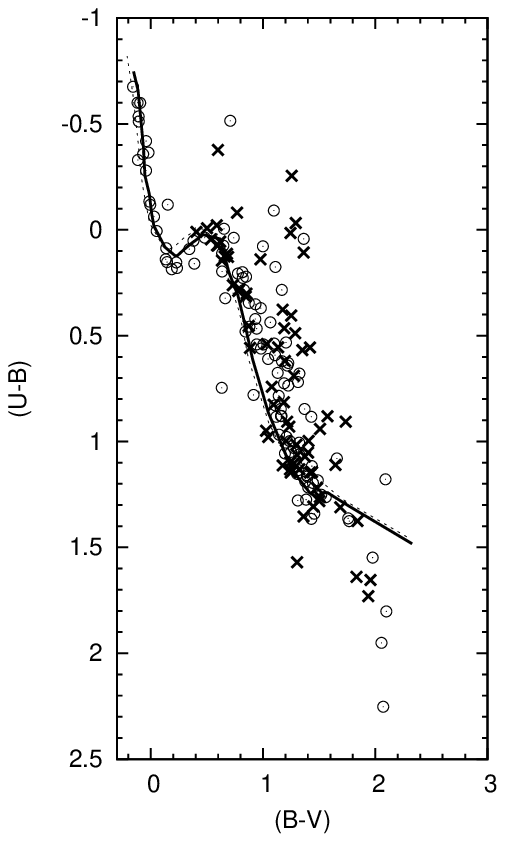}
\includegraphics[width=4.5cm,angle=0]{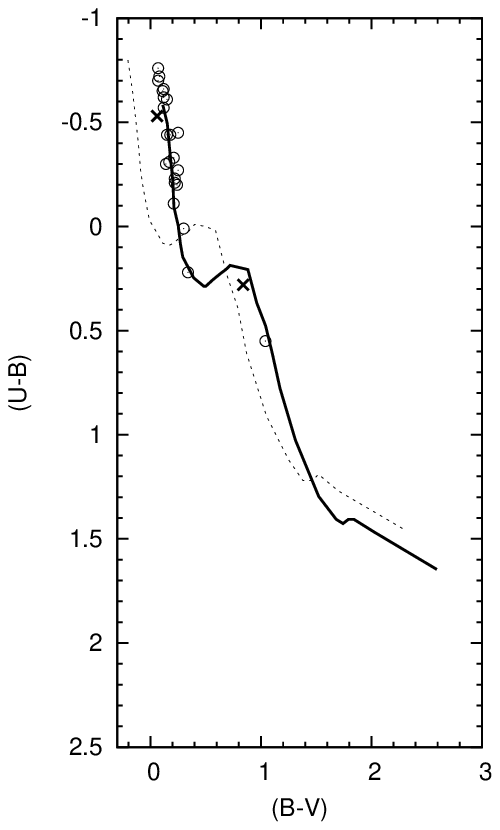}
\includegraphics[width=4.5cm,angle=0]{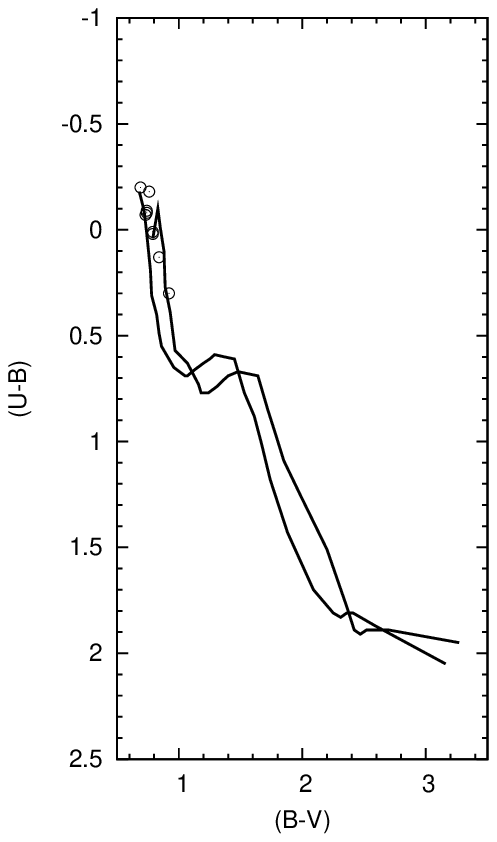}
\includegraphics[width=4.5cm,angle=0]{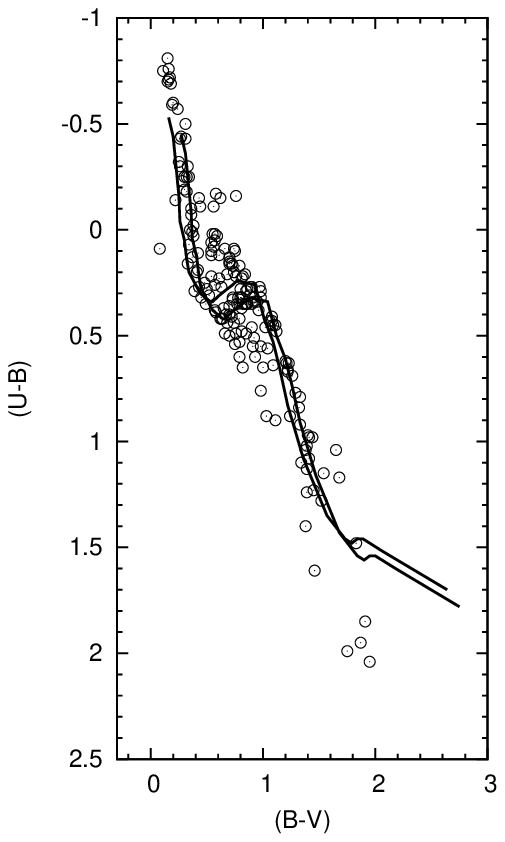}
\includegraphics[width=4.5cm,angle=0]{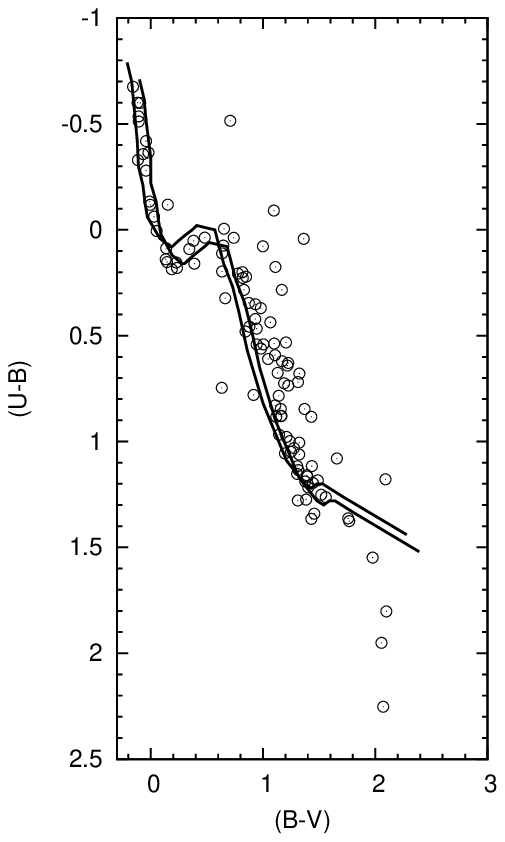}
\includegraphics[width=4.5cm,angle=0]{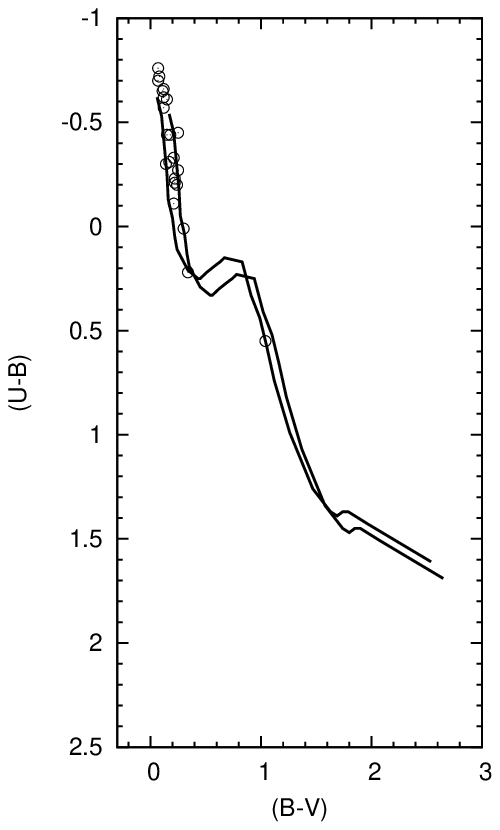}
\caption{Top: Reddened ZAMS fitting to the cluster members with P\%$>$P$_{min}$ (open circles). The position of the other stars of the sample 
is indicated by crosses. A full line shows the curve obtained with the  fitting algorithm, while the dotted line shows the intrinsic  U-B and  B-V colours. 
Bottom: Curves representing the limit $\Delta E(B-V)$ = 0.11 suggested by Burki (1975). 
From left to right:  Berkeley~86, NGC~2244, NGC~2264, and NGC~6530.}
\end{center}
\end{figure*}

\end{appendix}
\end{document}